\documentclass[letterpaper,twocolumn,10pt]{article}
\usepackage{usenix}
\usepackage{titling}
\usepackage{tikz}
\usepackage{multirow} 
\usepackage{pifont}
\usepackage{amsmath}
\usepackage{amssymb}
\usepackage{url}
\usepackage{hyperref}
\usepackage{subcaption}
\usepackage{graphicx}
\usepackage{booktabs}
\usepackage{algpseudocode}
\usepackage[linesnumbered,ruled,vlined]{algorithm2e}
\usepackage{tabularx}
\usepackage{authblk}
\usepackage{tablefootnote}
\usepackage{float}
\usepackage{afterpage}

\usepackage{enumitem} 
\usepackage{pifont}
\usepackage{placeins}
\newcommand{\mixed}{\ding{51}\kern-0.62em\ding{55}} 

\usepackage{tikz}
\newcommand*\circled[1]{%
  \scalebox{0.78}{\begin{tikzpicture}[baseline=-3pt]
    \node[draw,circle,inner sep=0.5pt, fill=black] {\textcolor{white}{\textsf{\textbf{#1}}}};
  \end{tikzpicture}}}

 \usepackage{etoolbox}
 
 \makeatletter
 \patchcmd{\maketitle}
 	{\@maketitle}
 	{\vspace{-7em}\@maketitle\vspace{-5em}}
 	{}
 	{}
 \makeatother

 \usepackage{listings}
\algdef{SE}[LOOP]{Loop}{EndLoop}[1]{\textbf{loop} #1}{}

\newcommand{\tinyskip}{\vspace{3pt}}
\newcommand{\mypar}[1]{\tinyskip\noindent\textbf{#1.}\xspace}

\makeatletter
\DeclareRobustCommand\onedot{\futurelet\@let@token\@onedot}
\def\@onedot{\ifx\@let@token.\else.\null\fi\xspace}

\makeatother

\title{\textbf{nncase: An End-to-End Compiler for Efficient LLM Deployment on Heterogeneous Storage Architectures}}

\author{
  Hui Guo\thanks{Email: \texttt{sunnycase@live.cn}}, 
  Qihang Zheng\thanks{Email: \texttt{zhengqihang0915@qq.com}}, 
  Chenghai Huo, 
  Dongliang Guo, 
  Haoqi Yang, 
  Yang Zhang
}

\affil{Canaan Inc.}

\date{} 

\begin{document}
\maketitle

\begin{abstract}

The efficient deployment of large language models (LLMs) is hindered by memory architecture heterogeneity, where traditional compilers suffer from fragmented workflows and high adaptation costs. We present nncase, an open-source, end-to-end compilation framework designed to unify optimization across diverse targets. Central to nncase is an e-graph-based term rewriting engine that mitigates the phase ordering problem, enabling global exploration of computation and data movement strategies. The framework integrates three key modules: Auto Vectorize for adapting to heterogeneous computing units, Auto Distribution for searching parallel strategies with cost-aware communication optimization, and Auto Schedule for maximizing on-chip cache locality. Furthermore, a buffer-aware Codegen phase ensures efficient kernel instantiation. Evaluations show that nncase outperforms mainstream frameworks like MLC LLM and Intel IPEX on Qwen3 series models and achieves performance comparable to the hand-optimized llama.cpp on CPUs, demonstrating the viability of automated compilation for high-performance LLM deployment. The source code is available at \url{https://github.com/kendryte/nncase}.

\end{abstract}

\section{Introduction}
\vspace{-0.3cm}

As large language models (LLMs) such as GPT-4\cite{gpt4} and Llama\cite{llama} scale beyond the trillion-parameter mark, they impose exponential demands on computation and storage. This surge presents a fundamental challenge: navigating the heterogeneity of memory architectures. Contemporary deployment landscapes are bifurcated into distributed systems—spanning multi-server clusters and specialized accelerators (e.g., NVIDIA H100 arrays, Cerebras WSE\cite{wse,wse3})—and single-node setups on standalone devices. Despite their topological differences, these architectures share critical constraints:
(1) Deep Memory Hierarchies: Ranging from KB-level SRAM/L1 caches to GB-level global HBM/DRAM.
(2) Heterogeneous computation Units: The coexistence of scalar, vector (e.g. AVX-512\cite{intel-xeonphi}) and matrix (e.g. Intel AMX\cite{intel-Xeon6}) units.
(3) The Memory Wall: With the memory bandwidth growing only 20\% annually versus a 100\% surge in compute power, data movement has become the primary bottleneck for LLM inference.

Efficient deployment requires orchestrating the synergy between memory and computation, specifically targeting hierarchical reuse and heterogeneous unit adaptation. However, conventional compilers often necessitate distinct optimization pipelines for uniform and non-uniform memory architectures. This dichotomy introduces prohibitive adaptation costs and creates technical barriers to global optimization.

To bridge this gap, we present nncase, an open-source end-to-end compilation framework designed to unify LLM deployment across heterogeneous memory architectures. The cornerstone of nncase is its "unified distributed compilation paradigm." By modeling all targets via a Non-Uniform Memory Access (NUMA) abstraction, nncase decouples the compilation workflow from physical topology, achieving a "compile once, adapt everywhere" capability. The framework is underpinned by an e-graph-based term rewriting engine that employs equality saturation. Unlike the greedy strategies of traditional deep learning compilers, our engine circumvents the phase ordering problem, exploring a comprehensive design space without compromising semantic integrity. Furthermore, by integrating a Roofline-based cost model, nncase performs multi-objective optimization to balance memory footprint, latency, and communication overhead.

Using the capacity of the e-graph to represent equivalence, nncase enables three distinct optimization passes:

\textbf{1) Auto Vectorize:} Addressing the challenge of mapping tensors to various hardware units, this module introduces MetaPackOperation and FoldNopPack rules. It simultaneously generates multiple packed layout candidates within the e-graph, dynamically adjusting packing factors to eliminate redundant layout permutations. This allows the compiler to strike an optimal balance between data layout conversion and computing unit saturation.

\textbf{2) Auto Distribution:} To manage data parallelism, we adopt the SBP (Split, Broadcast, Partial-value)\cite{oneflow} abstraction. Operating on the principle that "nodes with consistent SBP attributes are equivalent," this module embeds the distributed strategy search space directly into the e-graph. It automatically balances communication costs against computational efficiency, adapting seamlessly to both multi-device clusters and single-node logical distributions.

\textbf{3) Auto Schedule:} Addressing the complexity of on-chip memory management, we introduce the nncase Tensor Template (NTT) Library. This library abstracts hardware primitives into atomic microkernels (µkernels). We then decompose scheduling into structural optimization (e.g., Loop Fusion) through Monte Carlo Tree Search (MCTS)\cite{mcts} and parametric optimization (e.g., Tiling) through Mixed-Integer Nonlinear Programming (MINLP)\cite{branch-bounds-minlp}. This hierarchical approach achieves register-level efficiency while avoiding the prohibitive search times of exhaustive methods.

We evaluated nncase using the Qwen3\cite{qwen3} model family (0.6B and 1.7B) on an AMD Ryzen 9 5900X platform in single-core and multi-core configurations. Experimental results demonstrate that nncase significantly outperforms mainstream frameworks such as MLC LLM\cite{mlc-llm} and Intel IPEX\cite{ipex}, while matching or surpassing the performance of the hand-optimized library llama.cpp\cite{llama_cpp}.

In summary, the contributions of this paper include the following:
\begin{itemize}
\item We design and open-source \href{https://github.com/kendryte/nncase}{nncase}, a unified compilation framework that bridges the gap between uniform and non-uniform memory architectures.
\item We propose an e-graph-based Auto Vectorize mechanism that resolves tensorization challenges across diverse compute units.
\item We develop an Auto Distribution method that embeds distributed strategy search into the e-graph, enabling topology-agnostic parallelism.
\item We introduce a hierarchical Auto Schedule system backed by the NTT Library, automating the optimization of on-chip memory reuse and instruction throughput.
\end{itemize}

\section{Background and Motivation}\label{sec:motivation}

\subsection{Data Layout Optimization}

Data layout optimization remains a fundamental bottleneck for deep learning compilers targeting heterogeneous architectures. Modern processors have evolved from purely scalar designs to sophisticated hybrids integrating both Vector Processing Units (e.g., AVX-512, NEON\cite{intel-xeonphi,armv8}) and dedicated Tensor Processing Units (e.g., Intel AMX\cite{intel-Xeon6}, Arm SME\cite{arm_sme}, Apple AMX\cite{apple_amx}). To harness this heterogeneous compute power, industry optimization strategies generally fall into two paradigms: Kernel-Level Optimization and Graph-Level Layout Propagation.

\textbf{Kernel-Level Optimization.} This paradigm treats layout optimization as a local problem isolated to individual operators. Representative frameworks, including vendor libraries (e.g., Intel MKL\cite{intel_mkl}, Arm Compute Library\cite{arm_acl}) and traditional compilers (e.g., Halide\cite{halide}, TVM\cite{tvm}), focus on aligning data with hardware intrinsics. For element-wise operations, loop vectorization is employed to match SIMD lane width; for compute-intensive operators like GEMM or Convolution, Just-In-Time (JIT) packing mechanisms (pioneered by GotoBLAS\cite{gotoblas}) are used to mitigate cache associativity conflicts.
\textit{Critique:} While locally optimal, this approach suffers from layout thrashing. By failing to exploit layout reusability across the computation graph, it incurs redundant packing and unpacking overheads at operator boundaries, significantly degrading end-to-end inference latency.

\textbf{Graph-Level Layout Propagation.} To address the redundancy of local packing, recent deep learning compilers employ global analysis to coordinate data layouts across the graph. Frameworks like TPP-MLIR\cite{tpp_mlir} and ALT\cite{alt} utilize term rewriting systems to propagate preferred layouts between upstream and downstream nodes, fusing layout transformations via constant folding or dead code elimination.
\textit{Critique:} Although these methods reduce transformation overheads, they lack the granularity to navigate the Vector-Tensor trade-off in modern hybrid architectures. When Vector Units (requiring 1D layouts) and Tensor Units (requiring blocked 2D layouts) coexist, the optimal strategy is not static. Existing approaches often fail to model the complex cost trade-off between the acceleration gains of 2D tensorization and the overhead of 1D-to-2D layout permutation, leading to sub-optimal execution plans in mixed-precision or mixed-unit scenarios.

\subsection{Distributed Strategy Search}

The distributed strategy search is the cornerstone of unlocking the full potential of multi-core/multi-chip clusters. Its primary objective is to partition large-scale LLM models and map them onto hardware nodes, seeking a global optimal balance between computational efficiency, memory utilization, and communication overhead. However, existing distributed optimization frameworks face fundamental challenges in achieving this unified goal.

\textbf{Reliance on Manual Heuristics.} Frameworks like GSPMD and OneFlow\cite{oneflow} rely heavily on manual intervention. GSPMD necessitates user-annotated sharding for critical tensors and propagates these strategies via broadcast mechanisms. Although OneFlow supports automatic derivation, it still requires manual seeding for initial layers. These semi-automatic approaches incur high engineering costs and lack the flexibility to adapt to diverse hardware topologies without human guidance.

\textbf{Decoupled Search in Automatic Frameworks.} State-of-the-art automatic search frameworks, such as Alpa\cite{alpa} and AutoDDL\cite{autoddl}, typically employ a hierarchical optimization strategy. They decompose the problem into separate stages—optimizing inter-operator parallelism (pipeline) and intra-operator parallelism (tensor sharding) independently. Although this reduces search complexity, it reverses the strong dependencies between computation, memory, and network. For example, a strategy optimized solely for communication minimization might violate local memory capacity constraints. Crucially, these methods lack a unified representation to jointly optimize compute, memory, and network costs. Consequently, they often settle for local sub-optima, failing to find strategies that are globally efficient across all three dimensions in resource-constrained heterogeneous environments.

\subsection{Computation Kernel Scheduling} 

The performance of kernel scheduling is the decisive factor in determining the hardware utilization and throughput of compute cores. However, the inherent heterogeneity in instruction sets, on-chip memory hierarchies, and compute unit architectures across different platforms renders the "one-size-fits-all" scheduling strategy impractical. On the one hand, vendor-provided high-performance libraries (e.g., cuDNN, MKL) suffer from inflexibility, failing to cover the diverse and rapidly evolving fusion patterns of LLM operators. On the other hand, manual optimization requires profound architectural expertise and incurs prohibitive engineering costs, making it unscalable for cross-platform deployment.

To tackle these challenges, two categories of auto-scheduling approaches have been proposed. The first is \textit{learning-based scheduling}, represented by Ansor~\cite{ansor} and others~\cite{looper,looptune,roller,auto_halide,meta_schedule}. These methods employ random search and machine learning cost models to explore a vast space of loop optimizations (e.g., reordering, tiling, and fusion). Despite their flexibility, they suffer from excessive compilation times, which hinders the rapid deployment and iteration of LLMs. 
The second category is \textit{analytical modeling-based scheduling}~\cite{analytical_model,analytical_blis,chimera,tuna}, which determines parameters like tile sizes using static mathematical models. Although offering fast compilation, these methods are often limited to integer parameter tuning, lack support for complex operator fusion, and struggle with accurate performance estimation. 

Crucially, most existing auto-schedulers treat scalar computation as the atomic scheduling granularity. This abstraction gap makes it difficult to perform fine-grained register-level optimizations—such as precise control over loop unrolling factors to avoid register spills and maximize reuse—thereby failing to match the peak performance achievable by hand-tuned assembly kernels.

\subsection{Observations and Solutions}

\begin{figure*}[t]
    \centering
    \includegraphics[width=\textwidth]{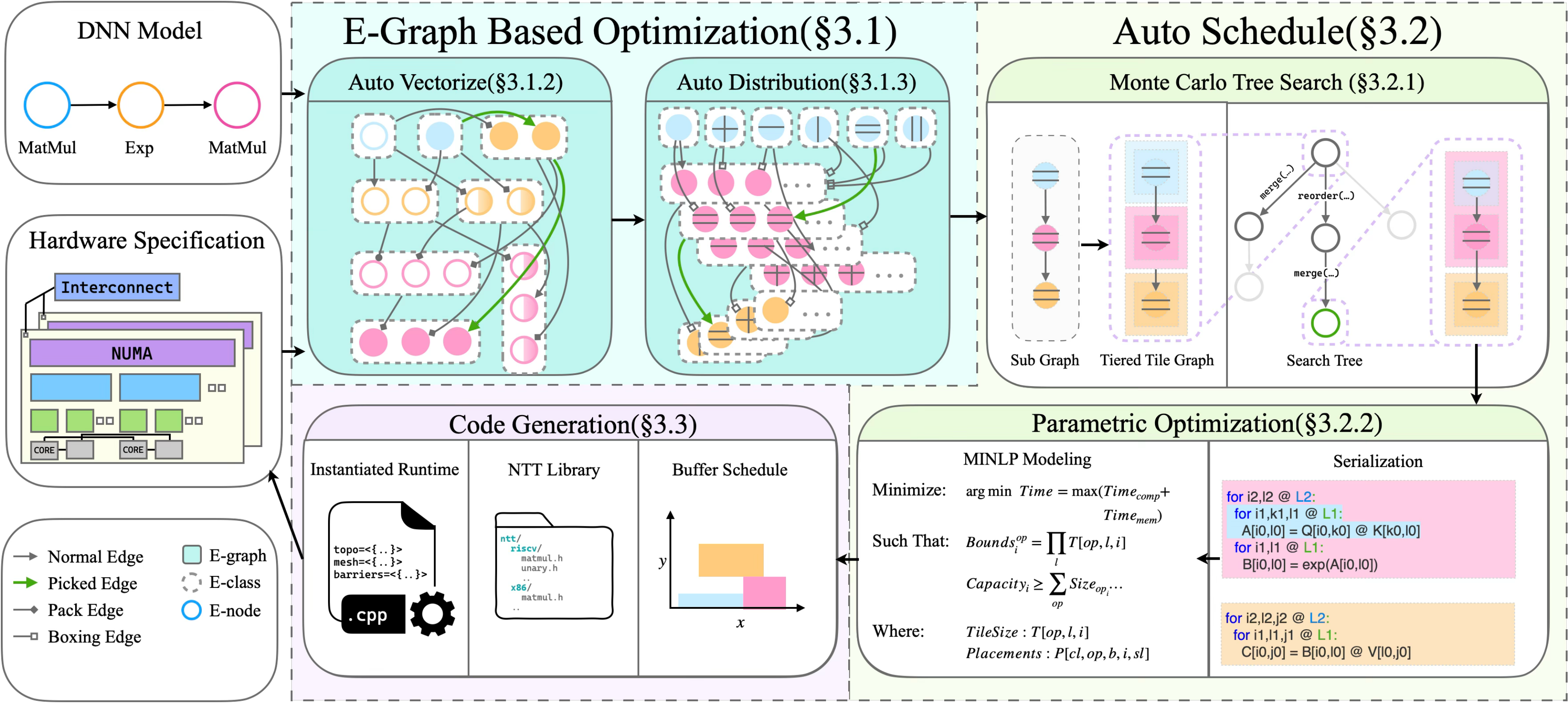}
    \caption{Overview of nncase.}
    \vspace{-0.5cm}
    \label{fig:arch:overview}
\end{figure*}

Motivated by the limitations of existing approaches and the deployment challenges identified above, we synthesize three key observations and propose the corresponding solutions implemented in \textit{nncase} (Figure~\ref{fig:arch:overview}):

\textbf{Observation 1: Layout Optimization Bottlenecks.} The primary obstacle in data layout optimization is the \textit{phase ordering problem} and the lack of global trade-offs. Traditional term rewriting engines perform destructive updates, irreversibly converting operators to a specific layout. This greedy approach prevents the exploration of alternative layouts (e.g., attempting a 2D pack after a 1D conversion) and lacks a quantitative comparison of conversion overheads versus computation gains.
\\
\textbf{Solution:} \textit{nncase} leverages \textbf{equality saturation} based on e-graphs to construct a non-destructive rewriting engine. We design a \textit{MetaPackOperation} rule that generates all candidate \textit{pack-compute-unpack} sequences in a single pass, preserving them within the e-graph. By evaluating the theoretical latency of each candidate using a Roofline-based cost model~\cite{roofline} and formulating the extraction as a constraint satisfaction problem (solved through a SAT solver~\cite{ortools}), we achieve global layout optimization that adapts to heterogeneous compute units.

\textbf{Observation 2: Suboptimal Distributed Strategies and Memory Risks.} Current distributed strategy searches suffer from two main defects: first, a decoupled cost modeling that optimizes communication while neglecting computation costs; second, the lack of rigorous memory capacity constraints. Existing schemes often prioritize throughput, leading to strategies that may trigger Out-Of-Memory (OOM) errors on resource-constrained devices, requiring tedious manual tuning to find a feasible configuration.
\\
\textbf{Solution:} We propose a formal description of distributed strategies based on the SBP abstraction~\cite{oneflow}. By defining "nodes with consistent SBP attributes" as equivalent, we map the distributed search space onto the e-graph structure. This allows us to reuse the e-graph's extraction mechanism to \textbf{jointly optimize computation and communication costs}. \textbf{Crucially, we formulate the strategy extraction as a constrained optimization problem solved by a SAT solver, where device memory capacity is enforced as a hard constraint,} ensuring that the automatically discovered strategy is not only performance-optimal but also strictly adheres to hardware memory limits without user intervention.

\textbf{Observation 3: Inefficient Kernel Scheduling.} Existing auto-schedulers struggle with a granularity mismatch: they schedule at the scalar level, missing register-level optimizations, while pure learning-based or analytical methods fail to balance code quality with compilation time.
\\
\textbf{Solution:} \textit{nncase} introduces the \textbf{Tensor Template Library (NTT)}, which encapsulates manually optimized register-efficient microkernels ($\mu$kernels) as atomic scheduling units. This hides low-level hardware complexity. On top of this, we decouple scheduling into two dimensions: \textit{structural optimization} (loop fusion/order) solved via Monte Carlo Tree Search (MCTS)~\cite{mcts}, and \textit{parametric optimization} (tile sizes/buffer location) solved via analytical modeling. This hybrid approach ensures register-level performance while significantly reducing search time compared to traditional methods.

\vspace{-0.3cm}
\section{Nncase Compiler Architecture}
\vspace{-0.2cm}

\autoref{fig:arch:overview} presents the overall architecture of nncase. The compilation pipeline is designed to bridge high-level model definitions with low-level hardware primitives through five distinct phases:

\begin{itemize}[leftmargin=*]
\item \textbf{E-Graph-Based Optimization (\S\ref{sec:arch:egraph_intro}:}
The workflow begins with ingestion of the computation graph in an e-graph structure (\circled{1}). This structure serves as the foundation for two critical tasks: determining the optimal data layout by term rewriting (\circled{2}, \S\ref{sec:arch:auto_vec}) and automatically searching for optimal distributed strategies by exploring splitting candidates and communication costs (\circled{3}, \S\ref{sec:arch:auto_dist}).

\item \textbf{Auto-Scheduling (\S\ref{sec:arch:auto_sche}):} 
For single-core subgraphs (\circled{4}), we convert the target subgraph into a tiered tile graph. We then employ a hybrid approach: Monte Carlo Tree Search (MCTS) optimizes the loop structure (\S\ref{sec:arch:mcts}), while a Mixed-Integer Nonlinear Programming (MINLP) solver optimizes tiling parameters based on a multi-memory hierarchy analytical model (\S\ref{sec:arch:sim}).

\item \textbf{Code Generation (\S\ref{sec:arch:codegen}):} 
In the final phase (\circled{5}), the compiler performs platform-specific memory scheduling and instantiates the runtime environment. The executable is generated by linking register-level high-performance microkernels (µkernels) provided by the NTT Library. This phase supports separate compilation flows to ensure maximum adaptation to specific hardware constraints.
\end{itemize}

\subsection{E-Graph Based Optimization}
\label{sec:arch:egraph_intro}

\subsubsection{Equality Saturation}
\label{sec:arch:equality_saturation}

\begin{figure*}[]
    \centering
    \includegraphics[width=\textwidth]{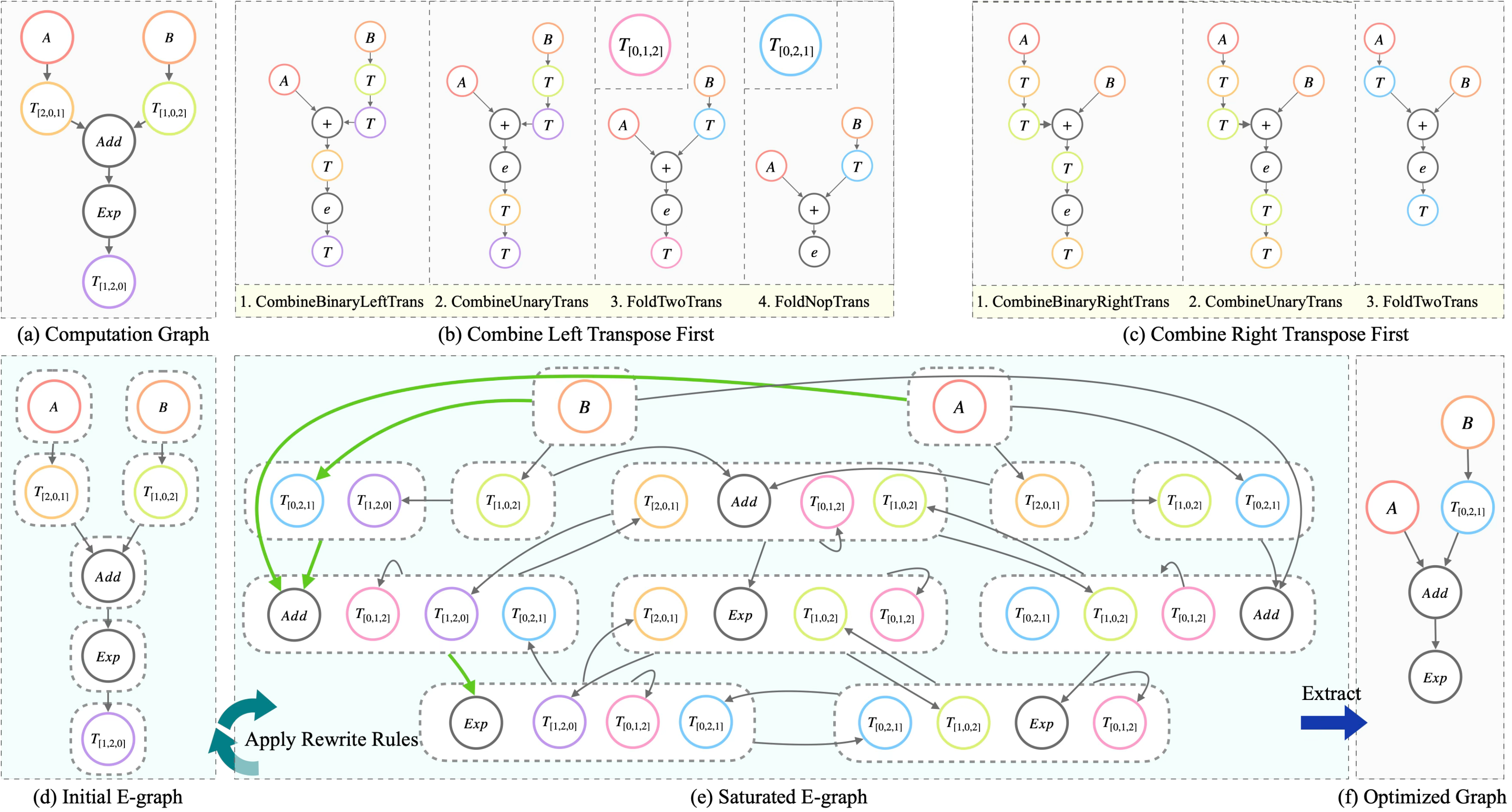}
    \caption{A comparison of traditional term rewriting and e-graph based rewriting.}
    \vspace{-0.5cm}
    \label{fig:arch:phase_order}
\end{figure*}

Term rewriting is a fundamental paradigm in program optimization, relying on the repeated application of simplification rules \cite{term_rewrite}. However, traditional sequential rewriting strategies suffer from the \textit{phase ordering problem} \cite{phase_order_1,phase_order_2}: the order in which rewrite rules are applied can destructively modify the graph, locking the optimization into a local optimum.

\begin{table}[t]
    \centering
    \resizebox{\linewidth}{!}{
        \begin{tabular}{c|c}
         \hline
         Rewrite Rule Name &	Signature  \\ 
         \hline
         CombineBinaryLeftTrans &	$Binary(T_{perm} (A),B)\rightarrow T_{perm} (Binary(A,T_{perm^{-1}} (B) ))$\\ 
         \hline
        CombineBinaryRightTrans &	$Binary(A,T_{perm} (B) )\rightarrow T_{perm} (Binary(T_{perm^{-1}} (A),B))$ \\ 
        \hline
        CombineUnaryTrans &	$Unary(T_{perm} (A))\rightarrow T_{perm} (Unary(A))$ \\ 
        \hline
        FoldTwoTrans &	$T_{perm_2} (T_{perm_1} (A))\rightarrow T_{perm_1 [perm_2[i]]} (A), 0 \leq i < Len(perm_2 )$ \\ 
        \hline
        FoldNopTrans	& $T_{\left[0,1,\ldots,n \right]} (A) \rightarrow A $  \\ 
        \hline
        \end{tabular}
        }
    \caption{Rewrite rules about transpose optimization.}
    \label{tab:arch:rules}
\end{table}

\autoref{fig:arch:phase_order} illustrates a typical example of this problem in deep learning compilers. Consider the computation graph in \autoref{fig:arch:phase_order}(a), where the objective is to eliminate redundant transpose operations using the rules defined in \autoref{tab:arch:rules}. The optimal strategy requires global reasoning about where to push the transpose operators.
\begin{itemize}
\item \textbf{Suboptimal Path (\autoref{fig:arch:phase_order}(c)):} If the compiler greedily applies \textit{CombineBinaryRightTrans} first, it isolates one transpose operation, preventing further fusion, and leaving a redundant operator in the final graph.
\item \textbf{Optimal Path (\autoref{fig:arch:phase_order}(b)):} Conversely, applying \textit{CombineBinaryLeftTrans} first, followed by \textit{CombineUnaryTrans}, enables the subsequent application of \textit{FoldTwoTrans} and \textit{FoldNopTrans}. This sequence successfully eliminates all redundant transposes.
\end{itemize}
This ambiguity highlights the limitation of greedy rewriting: it is often impossible to pre-determine the optimal rule application order without exploring the entire state space.

To resolve this, we adopt Equality Saturation \cite{equality_saturated_1,equality_saturated_2}, a non-destructive approach that explores the design space by embedding equality information into an e-graph structure. Instead of overwriting the IR, equality saturation applies all applicable rules simultaneously, growing the e-graph until it reaches a saturated state containing all equivalent versions of the program.

As depicted in \autoref{fig:arch:phase_order}(d), the e-graph consists of \textit{e-classes} (equivalence classes, shown as dashed boxes) and \textit{e-nodes} (computation nodes). An e-class represents a set of equivalent e-nodes, and each e-node references child e-classes rather than specific nodes. This compact structure allows efficient storage of exponential versions of the program.

Finally, extracting the optimal program from the saturated e-graph (\autoref{fig:arch:phase_order}(e)) is non-trivial. We formulate this extraction as a Weighted Partial MaxSAT (WPMAXSAT) problem \cite{egraph_extract_sat}. We utilize a Roofline-based cost model \cite{roofline}—incorporating metrics such as memory traffic and compute cycles—to assign weights to e-nodes. The solver then selects the subgraph with the minimal total cost, resulting in the fully optimized computation graph shown in \autoref{fig:arch:phase_order}(f).

\subsubsection{Auto Vectorize}
\label{sec:arch:auto_vec}

\textbf{Packing} (or Layout Transformation) is a fundamental prerequisite for utilizing heterogeneous computing units. It reorganizes tensors into hardware-intrinsic layouts—such as blocked formats for Tensor Cores or aligned arrays for Vector Units—to maximize spatial locality and instruction throughput. However, in heterogeneous graphs, distinct units often impose conflicting layout constraints. A naive compilation strategy typically results in redundant data movement, inserting costly \textit{Pack/Unpack} operations between operators with mismatched layout requirements.

To address this, we introduce an e-graph-based vectorization pass driven by two rewrite rules (see \autoref{tab:arch:pack_rules}):
\begin{enumerate}
    \item \textbf{MetaPackOperation (Exploration)}: Generates candidate packed subgraphs. For a given logical operator, it produces multiple hardware-specific variants (e.g., Blocked layouts for Tensor Cores vs. Flat layouts for Vector Units) wrapped in \textit{Pack/Unpack} nodes.
    \item \textbf{FoldNopPack (Optimization)}: Eliminates redundant layout transformations by fusing inverse operations ($Pack(Unpack(x)) \rightarrow x$).
\end{enumerate}

\begin{figure}[b]
    \centering
    \includegraphics[width=\linewidth]{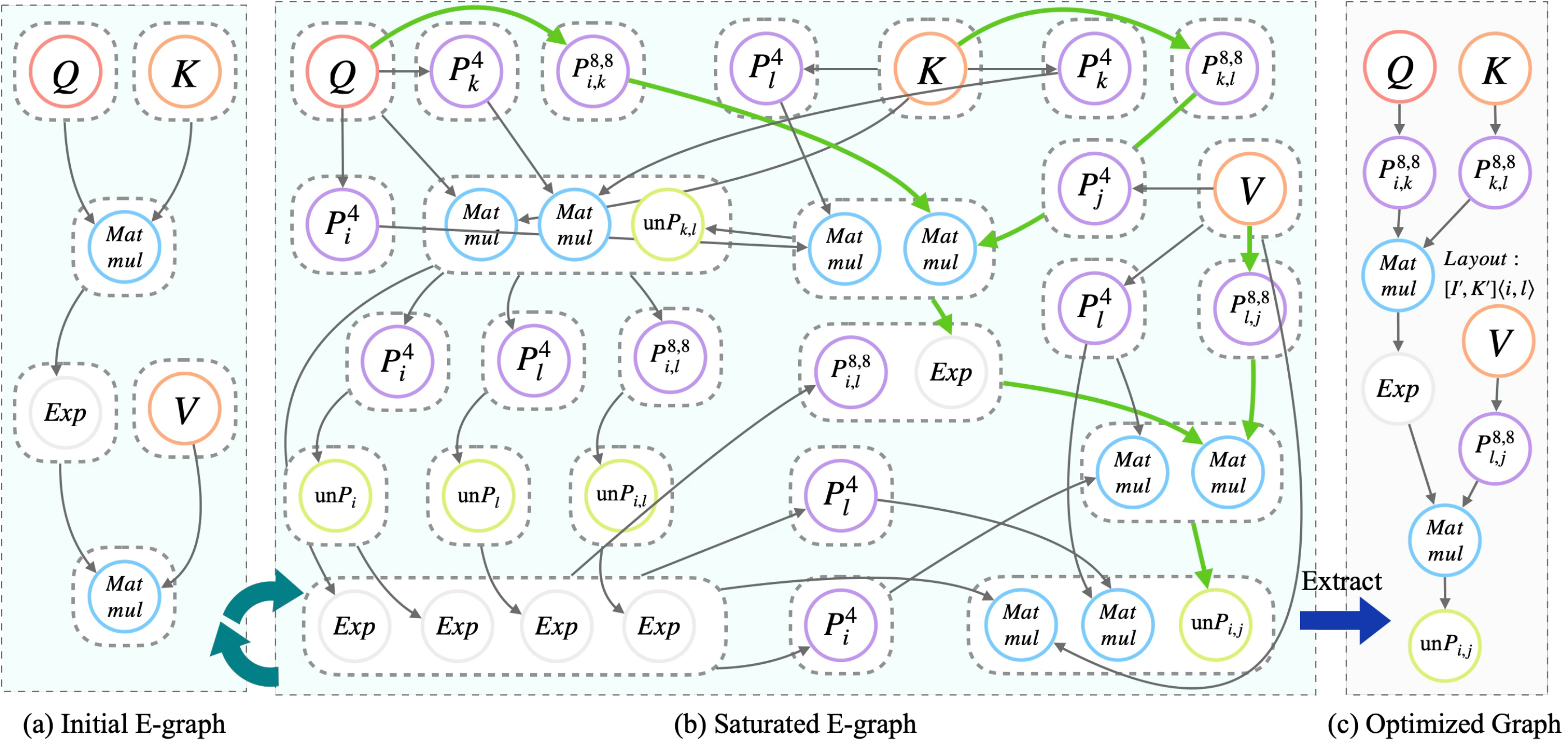}
    \caption{e-graph based auto vectorize for the Attention-like subgraph.}
    \label{fig:arch:auto_vec}
\end{figure}

\begin{table}[t]
    \centering
    \caption{Rewrite rules for vectorization and layout optimization.}
    \label{tab:arch:pack_rules}
    \resizebox{\linewidth}{!}{
        \begin{tabular}{c|l}
        \hline
         \textbf{Rule Name} & \textbf{Signature}  \\ \hline
         \multirow{4}{*}{MetaPackOperation}  &	
          $Op(\dots) \rightarrow Unpack(PackedOp($ \\ 
          & $\quad Pack(arg_1, lanes_1, axes_1),$ \\ 
          & $\quad \dots,$ \\
          & $\quad Pack(arg_n, lanes_n, axes_n)), axes^{-1})$\\ 
        \hline
        FoldNopPack &	$Pack(Unpack(arg, \dots), \dots) \rightarrow arg$ \\ 
        \hline
        \end{tabular}
    }
\end{table}

We demonstrate this mechanism using an Attention-like subgraph: $O = MatMul(Exp(MatMul(Q, K)), V)$, as shown in \autoref{fig:arch:auto_vec}. This structure represents a typical ``Compute-Memory-Compute'' pattern where layout conflicts frequently occur:

\begin{enumerate}
    \item \textbf{Layout Conflict}: The first $MatMul(Q,K)$ prefers a blocked layout (e.g. $[M', N']\langle 16, 16 \rangle$) for Tensor Core acceleration. The subsequent element-wise $Exp$ typically expects a flat layout for standard Vector Units. The final $MatMul(\dots, V)$ again requires a blocked layout.
    \item \textbf{Space Exploration}: \textit{MetaPackOperation} expands the $Exp$ operator. Crucially, it generates a variant that operates directly on the blocked layout (treated the $16 \times 16$ block as a contiguous vector of length 256).
    \item \textbf{Layout Propagation}: Since the output of the packed $MatMul$ matches the input of this blocked $Exp$, and the output of $Exp$ matches the input of the final $MatMul$, the e-graph connects these nodes.
    \item \textbf{Redundancy Elimination}: The \textit{FoldNopPack} rule identifies and removes the intermediate \textit{Unpack} (after the first $MatMul$) and \textit{Pack} (before the final $MatMul$).
\end{enumerate}

Mathematically, the optimized data flow avoids restoring the logical layout $[M, N]$:
\begin{equation}
\begin{aligned} 
\text{Step 1:} & \quad T_1[M', N']\langle 16,16 \rangle = \text{MatMul}_{\text{tensor}}(Q_{\text{packed}}, K_{\text{packed}}) \\
\text{Step 2:} & \quad T_2[M', N']\langle 16,16 \rangle = \text{Exp}_{\text{vec}}(T_1) \\
& \quad \quad \text{\textit{\footnotesize // Direct op on blocks}} \\
\text{Step 3:} & \quad O[M', D']\langle 16,16 \rangle = \text{MatMul}_{\text{tensor}}(T_2, V_{\text{packed}})
\end{aligned} 
\end{equation}
By enabling this ``pass-through'' layout, the compiler extracts a graph where data remain in the optimal hardware format throughout the entire chain, significantly reducing memory access overhead.

\subsubsection{Auto Distribution}
\label{sec:arch:auto_dist}

To facilitate adaptation to distributed architecture, we define the distributed strategy search space using the SBP (Split, Broadcast, Partial) abstraction natively supported by OneFlow~\cite{oneflow}. This section details the fundamentals of SBP abstraction and the search space construction logic (\S \ref{sec:arch:auto_dist_1}), followed by the methodology for the construction of the search space based on e-graph and the extraction of optimal strategies (\S \ref{sec:arch:auto_dist_2}).

\paragraph{SBP Abstraction and Search Space Definition}
\label{sec:arch:auto_dist_1}

\begin{figure}[t]
    \centering
    \includegraphics[width=\linewidth]{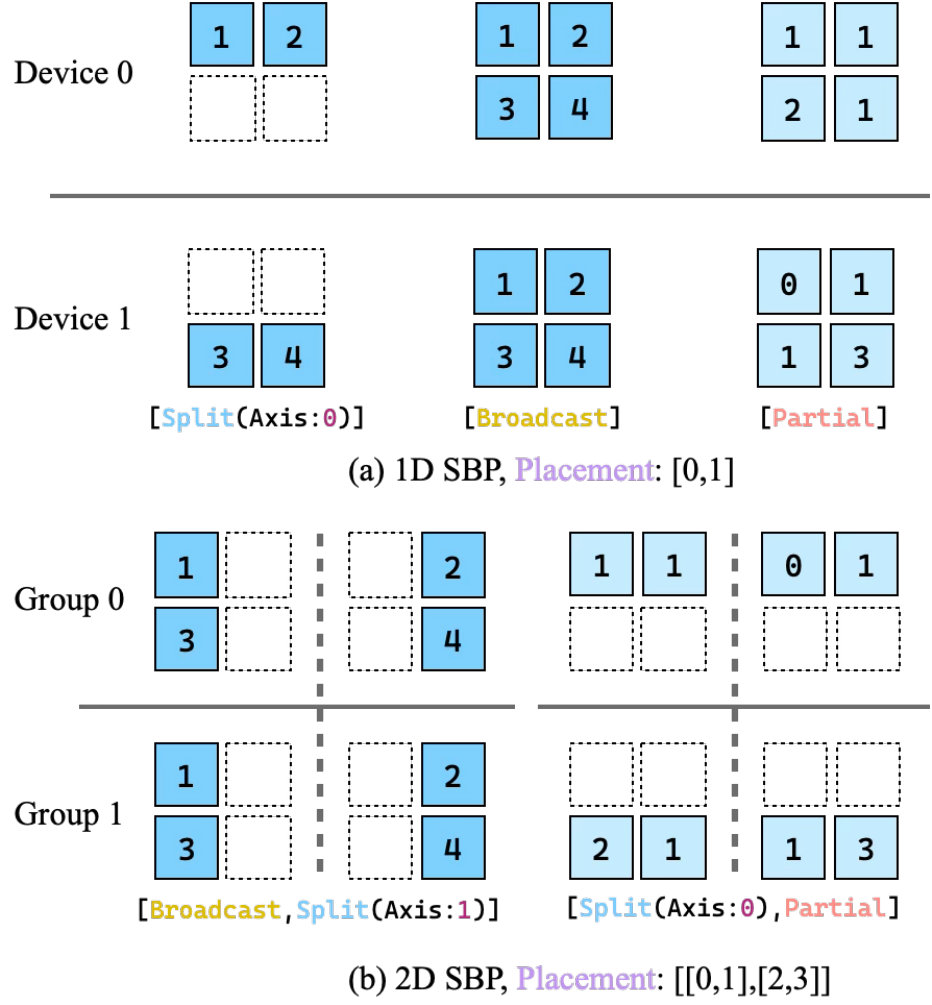}
    \caption{Illustration of SBP abstraction and device placement.}
    \vspace{-0.5cm}
    \label{fig:arch:sbp}
\end{figure}

The SBP abstraction accurately characterizes the distributed states of tensors across a device cluster. It describes how a $m$-dimensional tensor is distributed across an $n$-level logical topology using three primitives: \textit{Split}, \textit{Broadcast}, and \textit{Partial} (collectively called SBP), as illustrated in \autoref{fig:arch:sbp}. 
In addition to SBP, \textit{Placement} defines the logical topology of the device cluster. By mapping physical devices to a multi-dimensional array, the array's axes represent different topological levels. For example, \autoref{fig:arch:sbp}(b) depicts a nested topology comprising two groups of devices, each containing two devices.

To formalize the distributed state, the concept of N-Dimensional (ND) SBP is introduced to characterize the tensor distribution in a $N$-dimensional \textit{placement} . Specifically, ND SBP is defined as a vector of SBP operations, denoted as $\mathcal{P}_{\text{nd}} = [\text{sbp}_0, \text{sbp}_1, \ldots, \text{sbp}_{N-1}]$. 
As shown in \autoref{fig:arch:sbp}(a), the strategy for a $[2,2]$ tensor in a 1D \textit{placement} is defined by a 1D SBP:
\begin{itemize}
    \item \textbf{Split (S)}: Partitions the tensor along a specified axis.
    \item \textbf{Broadcast (B)}: Replicates the entire tensor to all devices.
    \item \textbf{Partial (P)}: Stores partial values (e.g., partial sums) on each device, requiring a reduction operation to recover the full data.
\end{itemize}
Crucially, SBP operations across different topological dimensions are orthogonal. As shown in \autoref{fig:arch:sbp}(b), a 2D SBP characterizes the strategy under a 2D topology, a paradigm extensible to higher dimensions. Thus, SBP abstraction offers a generalized capability to describe the tensor distribution under arbitrary valid topologies.

Although SBP effectively describes the state of individual tensors, the data flow in a computation graph imposes strict constraints on SBP configurations. The distributed attributes of upstream and downstream tensors cannot be arbitrarily assigned; they must strictly adhere to the operator's semantic requirements. 
Consider an element-wise binary addition $C = A + B$, where the inputs $A, B$ have shape $[4,4]$. On a two-device cluster, the validity of the distributed execution depends on the consistency of the operands' SBP states:

\begin{equation}
\label{eq:sbp_validity}
\begin{aligned}
    \text{Valid:} & \quad A_{\mathcal{S}(0)} + B_{\mathcal{S}(0)} \to C_{\mathcal{S}(0)} \\
    \text{Invalid:} & \quad A_{\mathcal{S}(0)} + B_{\mathcal{S}(1)} \to \bot 
\end{aligned}
\end{equation}

As demonstrated in Eq.~\ref{eq:sbp_validity}, combining a tensor split along the axis 0 ($\mathcal{S}(0)$) with a split along the axis 1 ($\mathcal{S}(1)$) for elemental addition is mathematically undefined without redistribution. 
We define the set of all valid input-output SBP combinations for an operator as its \textbf{SBP Signature}. The global search space for distributed strategies is thus constructed from the composition of these legal SBP Signatures across the entire network.

\paragraph{Search Space Construction and Extraction}
\label{sec:arch:auto_dist_2}

We construct the e-graph based on the principle that nodes sharing identical computation logic and SBP attributes are equivalent. While standard E-Graph rewriting rules could theoretically explore the space, adding distributed equivalents one by one incurs prohibitive compilation overhead. To address this, we propose a dedicated algorithm to directly construct a distributed E-Graph from the logical computation graph.

For a logical operator $Op(A)$, its distributed execution is formalized as $Box(Op(Box(A, \mathcal{P})))$, where \textit{Box} serves as a unified communication primitive. Boxing handles all necessary data movements—such as splitting, broadcasting, aggregation, and resharding—to satisfy SBP requirements. 
To manage the mapping between the original logical nodes and the distributed E-Nodes, we introduce the concept of an \textbf{E-Cluster}. An E-Cluster is a dictionary that maps a logical node to a set of E-Classes, grouped by their SBP signatures.

\begin{figure}[t]
    \begin{minipage}[t]{\linewidth}
    \begin{lstlisting}[language=Python, numbers=left, basicstyle=\scriptsize\ttfamily, escapechar=|, xleftmargin=1.5em, frame=single, breaklines=true]
def BuildEGraph(G):
    eg, memo = EGraph(), {} 

    # 1. Inputs: Create initial Shard Boxing
    for v in G.Inputs:
        # Generate base candidates
        nodes = [Box(v, s) for s in GetSBPs(v)]
        memo[v] = {}
        # GroupBySBP returns {sbp: [nodes]}
        for s, ns in GroupBySBP(nodes).items():
            memo[v][s] = eg.Add(ns)

    # 2. Compute: Topo-sort & Expansion
    for v in TopoSort(G):
        if v in G.Inputs: continue
        
        # Expand: Reuse + Resharding
        in_grps = []
        for inp in v.Inputs:
            # Reuse existing E-Classes
            cands = list(memo[inp].values()) 
            # Add new Resharding strategies
            cands += GenReshard(inp) 
            in_grps.append(cands)
        
        # Cartesian Product & Check
        nodes = []
        for args in Product(*in_grps):
            if IsValid(v, args):
                nodes.append(CreateOp(v, args))
        
        # Add to E-Graph & Union equivalents
        memo[v] = {}
        for s, ns in GroupBySBP(nodes).items():
            ids = [eg.Add(n) for n in ns]
            memo[v][s] = eg.Union(*ids)

    # 3. Outputs: Aggregate to Host
    for v in G.Outputs:
        # Unshard to Host
        nodes = [Box(n, Host) for n in memo[v].values()]
        ids = [eg.Add(n) for n in nodes]
        eg.Union(*ids)

    return eg
    \end{lstlisting}
  \end{minipage}
  \caption{Pseudo-code for the distributed strategy search space construction.}
  \label{alg:build_egraph}
\end{figure}

The construction process, detailed in \autoref{alg:build_egraph}, proceeds in three distinct phases:

\begin{enumerate}
    \item \textbf{Input Phase (Lines 5-11):} For each graph input, we generate all feasible distributed strategy candidates. These are instantiated as Boxing nodes and registered in the e-graph. The resulting e-classes are stored in the node's e-cluster.
    
    \item \textbf{Compute Phase (Lines 14-36):} This is the core of the search space expansion. Reusing simply the SBP states of the upstream nodes often leads to a shrinking search space. To mitigate this, we explicitly insert \textit{Resharding Boxing} operations (Line 23). 
    We generate input combinations via a Cartesian product of the inputs' e-Classes. Valid combinations are instantiated as compute nodes. Crucially, nodes resulting in the same output SBP are merged into a single e-class using the \texttt{Union} operation (Line 36), ensuring that the graph remains compact.
    
    \item \textbf{Output Phase (Lines 39-43):} Finally, to ensure that results can be retrieved by the host, we append \textit{Unshard Boxing} operations to output nodes, unifying all distributed branches into a global result.
\end{enumerate}

\begin{figure}[t]
    \centering
    \includegraphics[width=\linewidth]{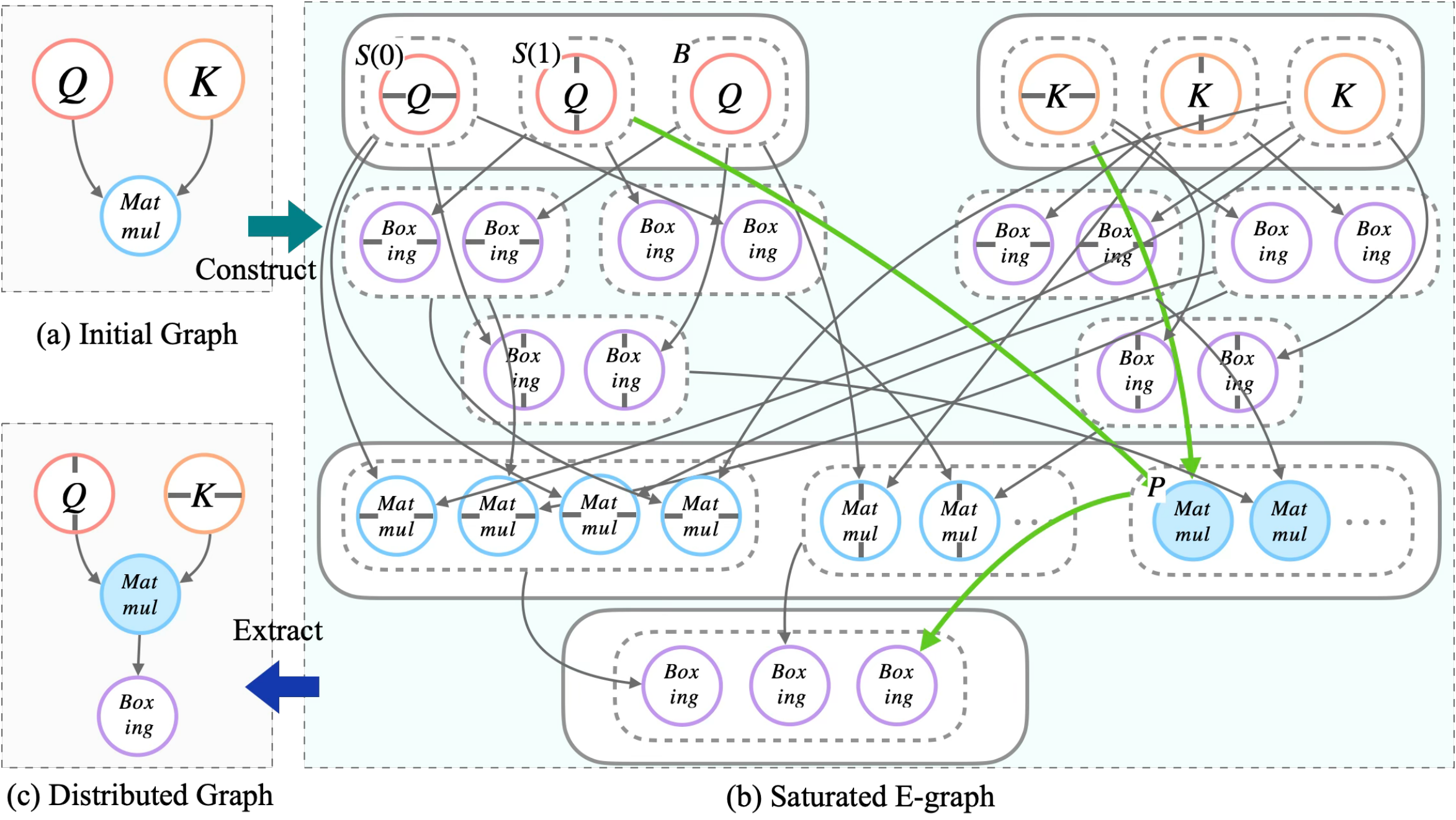}
    \caption{Visualization of the e-graph for distributed strategy search.}
    \vspace{-0.5cm}
    \label{fig:arch:auto_dist}
\end{figure}

\autoref{fig:arch:auto_dist}(b) visualizes this structure. Dashed boxes represent e-classes (equivalence classes of same SBP), while solid boxes represent e-clusters (collections of e-classes for one logical node). For example, the \textit{MatMul} node's e-cluster contains multiple e-classes derived from different valid SBP signatures. The search space is fully connected via Boxing nodes, allowing the extractor to find the optimal path.

\paragraph{Cost Model and Constraints}
To guide extraction, we incorporate a communication cost model based on the Roofline model and the Alpha-Beta model (latency-bandwidth) used in MPI implementations~\cite{mpich}. The simulated communication time serves as the edge cost in the e-graph.
Furthermore, to prevent the heuristic from favoring "Broadcast-heavy" strategies that minimize communication but explode memory usage, we enforce a memory constraint: a strategy is valid only if the sum of memory requirements across all devices does not exceed the cluster's capacity.

\subsection{Auto Scheduling}
\label{sec:arch:auto_sche}

\begin{figure}[t]
    \centering
    \includegraphics[width=\linewidth]{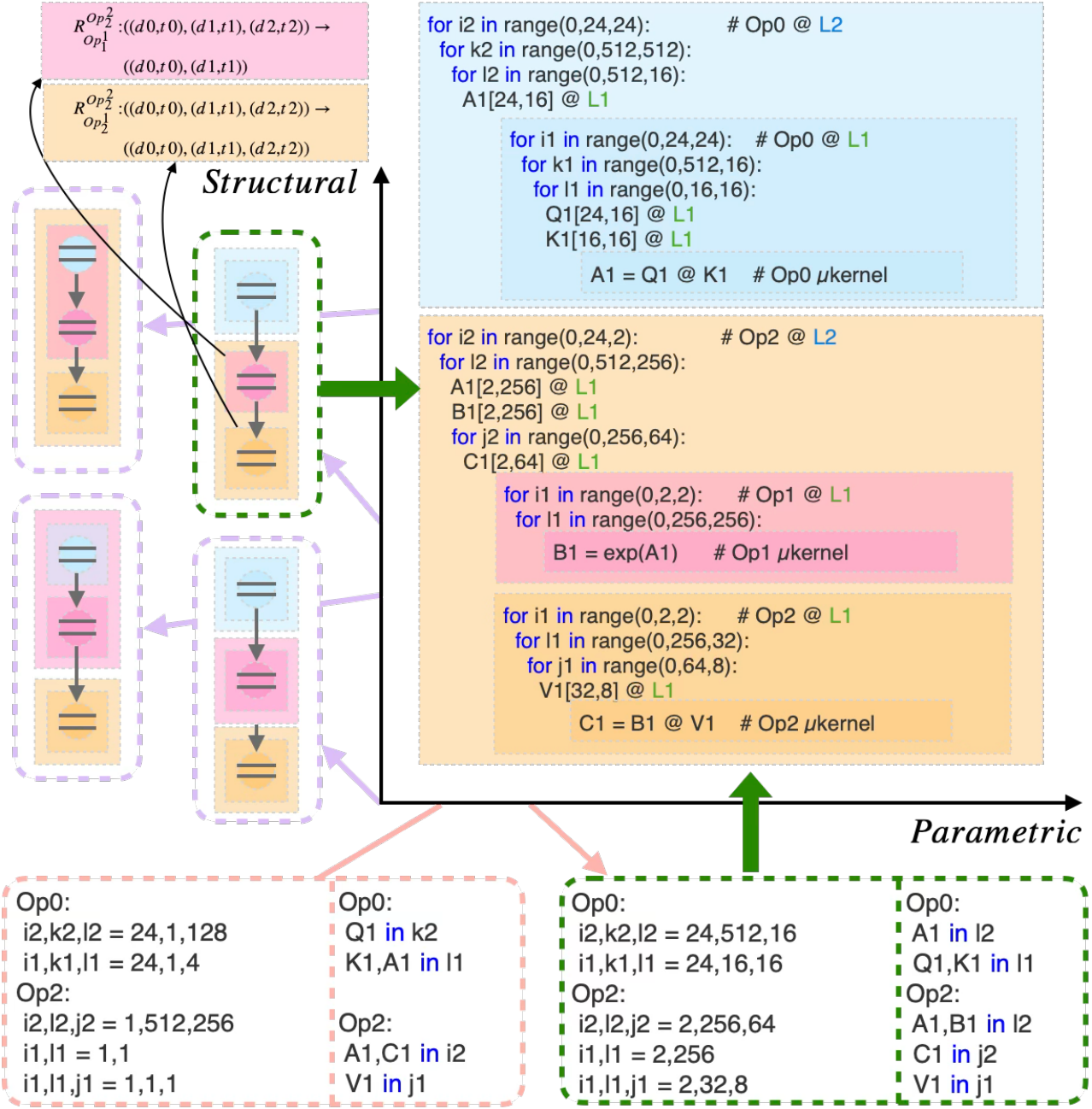}
    \caption{The design space of a computation kernel, decoupled into Structural and Parametric dimensions.}
    \vspace{-0.5cm}
    \label{fig:arch:auto_sche}
\end{figure}

We adopt tiles as the fundamental scheduling unit to construct high-performance computation kernels using stacking and nesting combinations~\cite{tileflow,welder,ladder,analytical_model}. As illustrated in \autoref{fig:arch:auto_sche}, we decouple the design space into two orthogonal dimensions: the \textbf{Structural Part} and the \textbf{Parametric Part}.

The \textbf{Structural Part} (y-axis) dictates the macroscopic organization of the kernel, including loop ordering, nesting levels, and parallelism. Each design point in this dimension corresponds to a unique \textit{Tiered Tile Graph}—a hierarchical data structure composed of nested Tile Graphs representing different memory levels.
This structure offers two key advantages:
\begin{itemize}
    \item \textbf{Dependency Representation:} It naturally characterizes computational dependencies derived from the original computation graph. The outer loops are mapped to the root nodes, while the inner loops are mapped to the child nodes, clearly depicting the hierarchy suitable for complex branching and merging scenarios.
    \item \textbf{Fusion Logic:} The nesting structure intuitively maps to operator fusion. As shown in the green dashed box in \autoref{fig:arch:auto_sche} (left), the Tile Graph for $Op_2$ contains two $L1$-level subgraphs. This indicates that $Op_2$ and $Op_1$ are fused at the $L2$ memory level. Specifically, $Op_2$ depends on $Op_1$, and intermediate results are transmitted exclusively within the $L2$ and inner memory levels, avoiding redundant off-chip memory access.
\end{itemize}

We attach polyhedral information to each Tile Graph to define the iteration domain mapping between the current node and its parent. For example, the affine map shown in the upper left of \autoref{fig:arch:auto_sche}, $R_{Op_1^1}^{Op_2^2}:((d0,t0),(d1,t1),(d2,t2) )\rightarrow((d0,t0),(d1,t1))$, implies that the $L1$-level loops of $Op_1$ are sliced from the first two dimensions of the parent $Op_2$.

The \textbf{Parametric Part} (x-axis) focuses on hardware-specific configurations, primarily \textit{Tile Size} and \textit{Buffer Placement}. These parameters directly impact data movement volume, cache utilization, and instruction efficiency.
\begin{itemize}
    \item \textbf{Tile Size:} Determines the granularity of data processing. Excessively large tiles cause cache thrashing due to capacity evictions, while overly small tiles lead to poor bandwidth utilization and cache fragmentation.
    \item \textbf{Buffer Placement:} Determining where the intermediate data resides. For a fixed tile size, allocating buffers in lower-level memory (when capacity allows) enables high-speed data reuse.
\end{itemize}

\autoref{fig:arch:auto_sche} (bottom) demonstrates the impact of these parameters. For the same structural graph, the red box configuration ($[1,1,1]$) suffers from poor locality. In contrast, the green box configuration ($[2,32,8]$) processes larger data blocks and allocates $A1$ and $B1$ buffers at the $L2$ level for $L1$ caching, significantly improving reuse efficiency.

\vspace{0.5em}
\noindent \textbf{Hybrid Search Strategy.}
Based on the orthogonality of these two dimensions, we propose a bi-level hybrid search framework to navigate this vast design space efficiently.
As detailed in the following subsections, we employ \textbf{Monte Carlo Tree Search (MCTS)} as the outer loop to explore the discrete structural space (\S \ref{sec:arch:mcts}), determining the optimal loop order and fusion strategy.
For each candidate structure visited by MCTS, we utilize \textbf{Mixed-Integer Nonlinear Programming (MINLP)} as the inner loop evaluation step (\S \ref{sec:arch:minlp}). This analytical solver rapidly identifies optimal tile sizes and buffer placements, providing a high-quality performance upper bound to guide the MCTS process.

\subsubsection{MCTS-based Structural Search}
\label{sec:arch:mcts}

The determination of the Structural Part involves defining the nested associations and loop execution orders within the Tiered Tile Graph. For a subgraph with $n$ operators, the combinatorial space of potential nested structures and loop permutations grows exponentially. We formulate this as a \textbf{sequential decision-making problem}: at each step, a transformation (e.g., fusion or reordering) is applied to the current graph state. These decisions are highly correlated—early structural choices constrain future optimization opportunities—making the problem analogous to a complex game tree search.

To navigate this high-dimensional and non-convex design space efficiently, we employ the \textbf{Monte Carlo Tree Search (MCTS)} algorithm. MCTS balances exploration and exploitation by constructing a search tree where nodes represent Tiered Tile Graph states, and edges represent transformation actions. Unlike greedy approaches, MCTS can look ahead to evaluate how current structural changes impact final performance.

\paragraph{Formal Representation}
We adopt the tile-centric notation~\cite{tileflow} to formally characterize the state of a Tiered Tile Graph. A node at the memory level $n$ is defined recursively as:
\begin{equation}
    Op^{n} = \{l_1^{n},l_2^{n},\ldots\}(Op^{n-1}_a, Op^{n-1}_b, \ldots)
\end{equation}
where $\{l_i^n\}$ denotes the loop set at level $n$, and $(Op^{n-1}_a, \ldots)$ denotes the topologically sorted list of child sub-graphs.
For instance, the Tiered Tile Graph in \autoref{fig:arch:auto_sche} is represented as:
\begin{align*} 
\text{Level 0: } & Op_0^0=\{\}(\text{MatMul}),\ Op_1^0=\{\}(\text{Exp}), \\
                 & Op_2^0=\{\}(\text{MatMul}) \\ 
\text{Level 1: } & Op_0^1=\{i^1,k^1,l^1\}(Op_0^0),\ Op_1^1=\{i^1,l^1\}(Op_1^0), \\
                 & Op_2^1=\{i^1,l^1,j^1\}(Op_2^0) \\ 
\text{Level 2: } & Op_0^2=\{i^2,k^2,l^2\}(Op_0^1), \\
                 & Op_2^2=\{i^2,l^2,j^2\}(Op_1^1,Op_2^1) 
\end{align*}

\paragraph{Search Mechanics}
The MCTS process iterates through four phases: \textit{Selection}, \textit{Expansion}, \textit{Simulation}, and \textit{Backpropagation}. The root node is initialized as the raw computation subgraph. During the \textit{Expansion} phase, we define two atomic actions to traverse the structural space:

\begin{itemize}
    \item \textbf{\texttt{merge(src, dst, level)}}: Implements \textit{operator fusion}. It migrates all subgraphs from the source node ($src$) to the destination node ($dst$) at a specified memory level, merging their loop scopes.
    Taking the operation \texttt{merge(1,2,2)} in the example above, the state transition at Level 2 is as follows:
    \begin{equation*}
        \begin{aligned}
            &Op_1^2=\{i^2,l^2\}(Op_1^1),\ Op_2^2=\{i^2,l^2,j^2\}(Op_2^1) \\
            &\xrightarrow{\text{\texttt{merge(1,2,2)}}} \quad Op_2^2=\{i^2,l^2,j^2\}(Op_1^1,Op_2^1)
        \end{aligned}
    \end{equation*}
    This action effectively fuses $Op_1$ and $Op_2$ at Level 2, as $Op_1^1$ becomes a child of $Op_2^2$.

    \item \textbf{\texttt{reorder(i, n, loops)}}: Implements \textit{loop permutation}. It reconfigures the execution order of the loop set for a specific tile $Op_i^n$.
\end{itemize}

\paragraph{Analytical Simulation}
A critical divergence from standard MCTS lies in the \textit{Simulation} phase. Instead of random rollouts, we employ a deterministic solver, specifically the \textit{Parametric Optimization via MINLP} module (detailed in \S\ref{sec:arch:minlp}), as the evaluator. For a given structural state (leaf node), this model solves for the optimal \textit{Parametric Part} (e.g., tile sizes) and returns the estimated latency as the reward. This hybrid approach—using MCTS for structural search and an analytical solver for parametric tuning—significantly reduces the search space and ensures high-quality evaluation signals without expensive hardware measurements.

\subsubsection{Parametric Optimization}
\label{sec:arch:minlp}

Once the Structural Part is determined, searching for the optimal parameter configuration is critical for execution performance. Performance is governed by complex interactions between factors such as cache capacity, buffer size, and tile size. Relying on empirical tuning is inefficient and unreliable. To systematically address this, we adopt an analytical modeling approach using \textbf{mixed integer nonlinear programming (MINLP)}.

MINLP is selected for two reasons: (1) it naturally handles integer variables (e.g., tile sizes) and nonlinear relationships inherent in hardware execution; (2) it supports complex constraints (e.g., memory limits) and multi-objective optimization. Compared to other methods, MINLP offers a flexible framework to accurately capture the "variable abstraction - static analysis - constraint definition - objective optimization" workflow.

\mypar{Variable Definition}
The variables represent the tuning parameters and are the primary subjects of the solution process.

\begin{enumerate}[label=(\arabic*), leftmargin=12pt, itemsep=2pt, topsep=2pt]

\item \textbf{Tile Size}: Each loop within a tiled code segment has an independent tile size. We define $\mathsf{TileVar}$ as a continuous integer variable:
\begin{equation}
    \mathsf{TileVar}_{l, op, i} \in \mathbb{Z}^+
\end{equation}
where $l$ denotes the memory level, $op$ the operator index, and $i$ the loop index.

\item \textbf{Buffer Placement}: We decouple buffer creation from storage to enable data reuse in faster caches. A boolean variable $\mathsf{Place}$ characterizes this:
\begin{equation}
    \mathsf{Place}_{cl, op, b, e, sl} \in \{0, 1\}, \quad 0 \leq sl \leq cl
\end{equation}
Here, $cl$ is the creation level, $b$ is the buffer ID, $e$ is the placement entry (position relative to loops), and $sl$ is the actual storage level. We restrict storage to levels at or below the creation level.

\end{enumerate}

\mypar{Static Analysis}
Performance metrics are derived by traversing the Tiered Tile Graph based on the defined variables.

\begin{enumerate}[label=(\arabic*), leftmargin=12pt, itemsep=2pt, topsep=2pt]

\item \textbf{Backward Extent}: While the Tile Graph records polyhedral information, the actual iteration domain depends on tile sizes. We define $\mathsf{Extent}$ recursively:
\begin{equation}
\begin{aligned}
    &\mathsf{Extent}_{l,op,e,i} = \mathsf{Extent}_{l-1,cop,0,i} \times \delta \\
    &\delta = \begin{cases} 
        \mathsf{TileVar}_{l,op,i}, & \text{if } i \geq e \land \mathcal{R}(loop^{l}, loop^{l-1}) \\  
        1, & \text{otherwise}
    \end{cases}
\end{aligned}
\end{equation}
where $\mathcal{R}$ determines if loops at adjacent levels map to the same iteration variable.

\item \textbf{Buffer Size}: The size of buffer $b$ is determined by its access relation $\mathcal{A}^b_{op}$ applied to the extent:
\begin{equation}
    \mathsf{Size}_{l,op,b,e} = \prod \mathcal{A}^{b}_{op}(\mathsf{Extent}_{l,op,b,e})
\end{equation}

\item \textbf{Trip Count}: The total number of iterations is the product of tile sizes from the current level down to the bottom:
\begin{equation}
    \mathsf{Trip}_{l,op,e} = \begin{cases}  
        \mathsf{Trip}_{l+1,op,last} & e=0 \\  
        \mathsf{Trip}_{l,op,e+1} \times \mathsf{TileVar}_{l,op,e} & e > 0  
    \end{cases}
\end{equation}

\item \textbf{Data Traffic}: We categorize buffers into read-only ($\mathcal{B}_{read}$) and read-write ($\mathcal{B}_{write}$) sets. The total data volume at memory level $l$ is:
\begin{equation}
\resizebox{0.9\hsize}{!}{$
\begin{aligned}
    \mathsf{Writes}_l &= \sum_{cl, op, e} \sum_{b \in \mathcal{B}_{write}} \Phi(cl, op, b, e, l) \\
    \mathsf{Reads}_l &= \sum_{cl, op, e} \sum_{b \in \mathcal{B}_{read}} \Phi(cl, op, b, e, l) \\
    \text{where } \Phi &= \mathsf{Place}_{cl,op,b,e,l} \times \mathsf{Size}_{l,op,b,e} \times \mathsf{Trip}_{l,op,e}
\end{aligned}
$}
\end{equation}
This model approximates data updates by multiplying loop counts with buffer sizes, avoiding the complexity of partial update analysis.

\end{enumerate}

\mypar{Constraints}
Constraints ensure the legality of the solution.

\begin{enumerate}[label=(\arabic*), leftmargin=12pt, itemsep=2pt, topsep=2pt]

\item \textbf{Domain Bounds}: The tiled loops must cover the original iteration domain. We enforce this by checking the shape at the top level ($L_{top}$):
\begin{equation}
    \mathcal{A}^{b}_{op}(\mathsf{Extent}_{L_{top},op,b,0}) = \mathsf{Shape}_{op}^{b}
\end{equation}

\item \textbf{Buffer Placement}: 
(a) I/O buffers must reside in the outermost memory:
\begin{equation}
    \mathsf{Place}_{L_{top},op,b,0,L_{top}} = 1
\end{equation}
(b) The innermost cache (level 0) must store the active buffer for computation:
\begin{equation}
    \sum_{cl, op, e} \mathsf{Place}_{cl,op,b,e,0} = 1
\end{equation}
(c) \textit{Unique Storage}: A buffer is stored exactly once per valid level.
(d) \textit{Fusion}: Fused operators must reuse buffers at the fusion level ($L_{fuse}$). For a reused buffer $b$:
\begin{equation}
    \sum_{sl=0}^{L_{fuse}}\sum_{e} \mathsf{Place}_{L_{fuse},op,b,e,sl} = 1
\end{equation}
Child nodes sharing this buffer must respect its dynamic storage level.

\item \textbf{Memory Capacity}: Total buffer usage at any storage level $sl$ must not exceed hardware capacity $C_{sl}$:
\begin{equation}
    \sum_{cl,op,b,e} \mathsf{Place}_{cl,op,b,e,sl} \times \mathsf{Size}_{cl,op,b,e} \leq C_{sl}
\end{equation}

\end{enumerate}

\mypar{Objective Function}
The goal is to minimize total execution time, modeled as the maximum of memory and computation time (assuming parallelism).

The computation time ($T_{comp}$) is derived from a linear regression model ($\mu\text{KernelTime}$) of the micro-kernel, scaled by iteration counts. Memory time ($T_{mem}$) is the data traffic divided by bandwidth ($BW_l$).
\begin{equation}
\begin{aligned}
    T_{comp} &= \sum_{op, b} \mu\text{KT}(op, \dots) \times \mathsf{Trip}_{0,op,last} \\
    T_{mem} &= \sum_{l} \frac{\mathsf{Writes}_l + \mathsf{Reads}_l}{BW_l}
\end{aligned}
\end{equation}

The final objective function solved by OR-Tools is:
\begin{equation}
    \text{minimize} \quad \max(T_{\text{mem}}, T_{\text{comp}})
\end{equation}

\subsection{Code Generation}
\label{sec:arch:codegen}

\begin{figure}[t]
    \centering
    \hspace{-0.5cm}
    \includegraphics[width=1.0\linewidth]{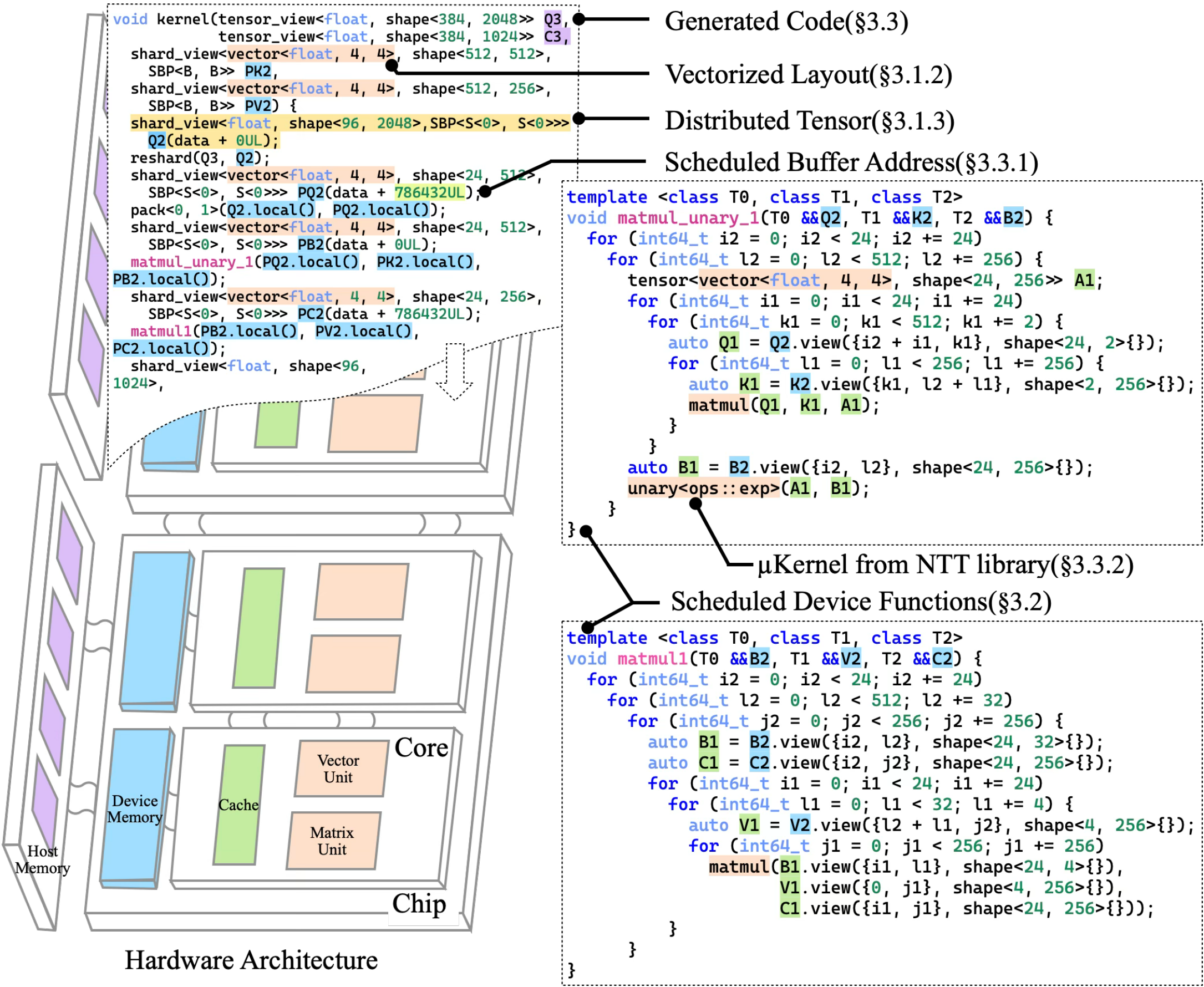}
    \caption{Generated C++ code utilizing NTT Library primitives.}
    \vspace{-0.3cm}
    \label{fig:arch:codegen}
\end{figure}

The code generation phase serves as the pivotal bridge between the front-end graph optimizations and the final executable binary. Its primary objective is to translate the optimized computation graph into efficient, hardware-aware C++ source code. As illustrated in \autoref{fig:arch:codegen}, the generation process relies on two critical components to maximize runtime performance:

\begin{itemize}
    \item \textbf{Buffer Scheduling}: This module addresses memory hierarchy constraints by optimizing allocation strategies and maximizing buffer reuse.
    \item \textbf{Nncase Tensor Template Library (NTT)}: This component facilitates the instantiation of the kernel template, leveraging high-performance register-level \textmu kernels to inject underlying hardware support into the generated code.
\end{itemize}

The following subsections detail the specific implementations of these two mechanisms.

\subsubsection{Buffer Schedule}

The core objective of Buffer Schedule is to minimize runtime memory redundancy and access overhead. This is achieved through a two-stage process: \textbf{Bufferization} (logical-to-physical mapping) and \textbf{Memory Planning} (address allocation).

\paragraph{Bufferization and Alias Analysis}
Before allocation, the compiler transforms the logical tensors into physical buffer constraints. A key optimization in this stage is \textbf{Alias Analysis}. The compiler identifies operators with view semantics, such as \texttt{Reshape}, \texttt{Slice}, and \texttt{Squeeze}, where the output tensor is merely a different view of the input data. Instead of allocating new storage, these outputs are marked as aliases, sharing the underlying physical memory of their inputs. This strategy enforces \textit{zero-copy} execution for shape transformations, significantly reducing memory footprint.

\paragraph{Memory Planning}
Once physical buffers are defined, the system assigns optimal memory addresses:
\begin{itemize}
    \item \textbf{Intermediate Variables:} By traversing the computation graph to perform liveness analysis, the system obtains the lifecycle and size of each non-aliased buffer. The allocation is modeled as a classic Bin Packing problem \cite{binpack,binpack2,opt_book}. An SAT solver is utilized to find an optimal arrangement, maximizing memory reuse by overlapping buffers that are not live simultaneously.
    \item \textbf{Constants:} For static weights, the system reuses the SBP (Split-Broadcast-Partial) attributes determined by the Auto Distribution module. Constants are pre-split and pinned to the dedicated local storage of each computing node. This pre-allocation strategy ensures high-speed local access and eliminates redundant data transmission in distributed scenarios.
\end{itemize}

\subsubsection{Nncase Tensor Template Library}

The nncase Tensor Template Library (NTT) serves as the foundational infrastructure for generated code, implemented as a header-only \textbf{C++20 template library}. Unlike traditional runtime operator libraries, NTT leverages modern C++ Metaprogramming (TMP) and Concepts to minimize runtime overhead through \textbf{Zero-cost Abstraction}.

\paragraph{Hybrid Shape and Stride System}
A distinguishing feature of NTT is its hybrid management of tensor metadata. As demonstrated by the usage of \texttt{ntt::fixed\_shape\_v} and \texttt{ntt::make\_shape}, the library allows mixing static compile-time dimensions with dynamic runtime dimensions.
\begin{itemize}
    \item \textbf{Static Inference:} For dimensions known at compile-time (e.g., model weights or fixed-size buffers), shape inference and stride calculations are performed entirely during the C++ compilation phase. This allows the compiler to resolve the offset calculations into immediate values, eliminating runtime arithmetic instructions.
    \item \textbf{Dynamic Compatibility:} For variable dimensions, the library seamlessly falls back to runtime computation, ensuring flexibility without compromising the performance of static parts.
\end{itemize}

\paragraph{Micro-kernels and Vectorization}
NTT provides a suite of architecture-aware micro-kernels tailored for core operators (e.g., MatMul, Conv2D). By exposing explicit vector types (e.g., \texttt{ntt::vector<float, 8>}) and memory layout primitives (e.g., \texttt{pack/unpack}), NTT enables the Codegen module to precisely control data reuse patterns and SIMD instruction generation. This design effectively avoids cache invalidation and guides the backend compiler (GCC/Clang) to generate optimal register spill/fill sequences.

\paragraph{Distributed Extension (NTTD)}
To support multi-device execution, the Distributed Tensor Template Library (NTTD) extends NTT with topology-aware abstractions.
\begin{itemize}
    \item \textbf{Type-Based Sharding:} Distributed attributes such as Mesh topology and SBP (Split-Broadcast-Partial) policies are encoded directly into C++ template arguments (e.g., \texttt{shard\_policy::S<2>}). This ensures that communication primitives are statically bound, avoiding runtime policy dispatch overhead.
    \item \textbf{Seamless Integration:} NTTD acts as the executable backend for the Auto Distribution module. The abstract tensor splitting strategies are mapped directly to NTTD's template configurations, enabling efficient overlapping of computation and communication on specific hardware topologies (e.g., NUMA, Chiplet clusters).
\end{itemize}

\section{Evaluation}
\label{sec:evaluation}
\vspace{-0.2cm}

This section evaluates the end-to-end inference performance of Large Language Models (LLMs) on a representative CPU platform using \textit{nncase}. The experiments aim to verify its advantages over mainstream compilation frameworks and validate the effectiveness of the proposed ``unified distributed compilation for heterogeneous storage architecture'' design.

We employ the Qwen3\cite{qwen3} series models, specifically Qwen3-0.6B (supporting the precision F32/F16) and Qwen3-1.7B (the precision F16). The evaluation was conducted on an AMD Ryzen 9 5900X 12-core processor equipped with 128GB ($4\times32$GB) DDR4-3600 memory, running Ubuntu 24.04 LTS (GCC 14.2). To ensure a fair comparison, all frameworks were configured with default optimizations and utilized the highest supported instruction set (AVX2).

We compare \textit{nncase}\footnote{Commit: fbe2f7ba9} against three state-of-the-art frameworks: MLC LLM\cite{mlc-llm}\footnote{Commit: 862a7311}, llama.cpp\cite{llama_cpp}\footnote{Tag: b5753}, and Intel IPEX\cite{ipex}\footnote{Version: v2.8.0}. Performance was evaluated under varying CPU thread configurations (1T, 4T, 8T) to assess scalability in both single-core and multi-core scenarios. Following industry standards, we set the batch size to 1 with an 8-token prompt. The primary metric is \textit{token throughput} (tokens/s), calculated based on the total duration of the decoding stage. Each test was repeated 100 times to report average values. \autoref{fig:eval:single_core} and \autoref{fig:eval:multi_core} illustrate the performance comparisons in different precisions and scales. Detailed analyzes are provided in the following subsections.

\vspace{-0.2cm}
\subsection{Single-Core Scenario}
\label{subsec:single_core}

This subsection evaluates the fundamental code generation quality of \textit{nncase} using Qwen3-0.6B (F32/F16) and Qwen3-1.7B (F16) in a single-thread (1T) configuration. 
Although \textit{nncase} employs a ``unified distributed compilation'' architecture, it treats a single core as the atomic unit of its distributed abstraction. Consequently, the performance in this scenario directly reflects the efficacy of the \textbf{Auto-Vectorization} and \textbf{Auto-Scheduling} modules, where the upper bound is determined by the compiler's ability to maximize on-chip memory reuse and compute unit utilization (e.g., AVX2 saturation).

\begin{figure}[t]
    \centering
    \includegraphics[width=\linewidth]{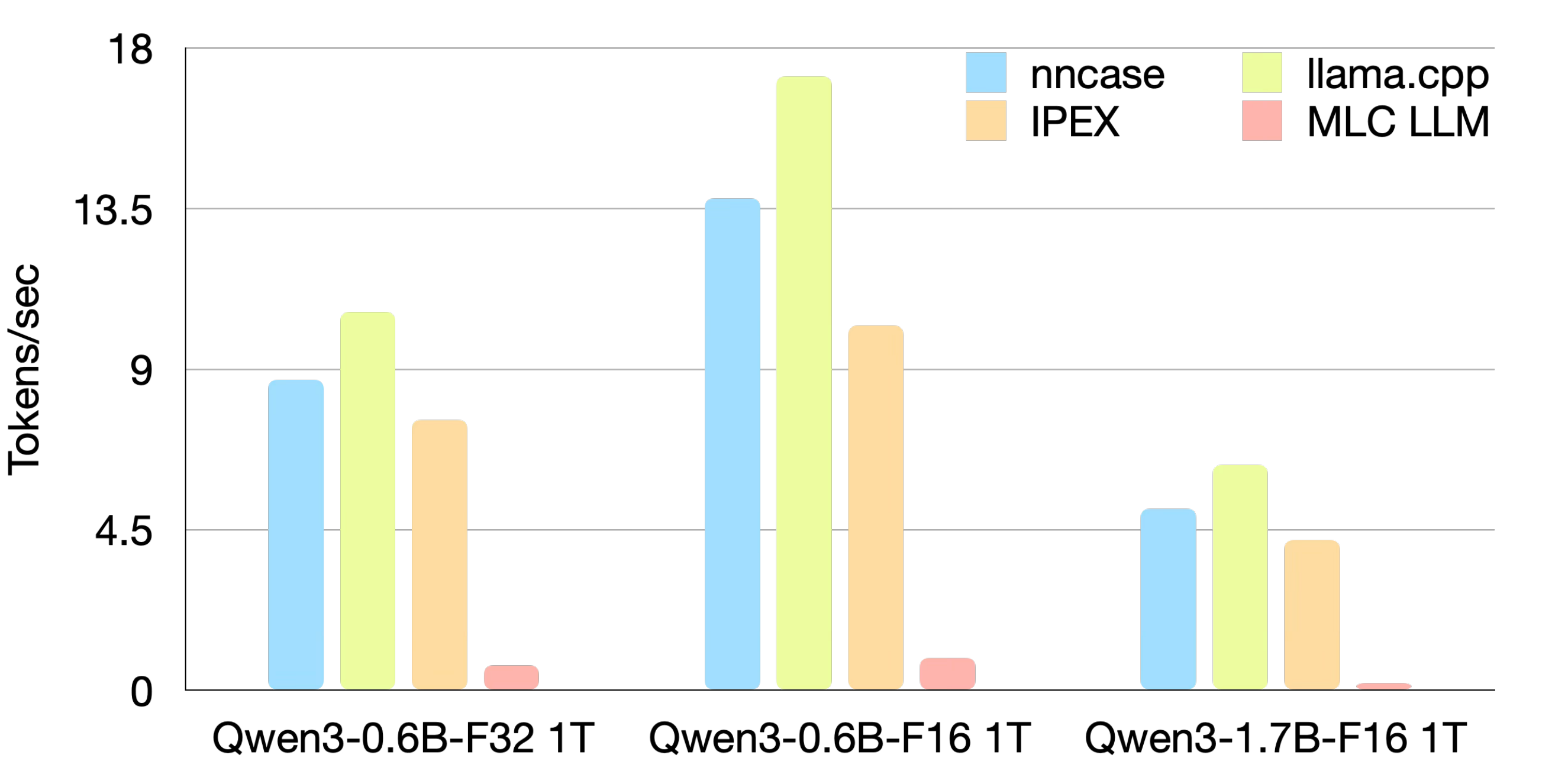}
    \caption{Comparison of LLM token generation throughput among different frameworks in the single-core scenario (1T). \textit{llama.cpp} represents the hand-optimized baseline, while others represent compiler-based approaches.}
    \vspace{-0.3cm}
    \label{fig:eval:single_core}
\end{figure}

As illustrated in \autoref{fig:eval:single_core}, the results exhibit a consistent performance hierarchy across different model scales and precisions: \textit{llama.cpp} $>$ \textit{nncase} $>$ \textit{Intel IPEX} $\gg$ \textit{MLC LLM}. 
\textit{llama.cpp} maintains the lead due to its extensive hand-written kernels and manual scheduling, serving as the theoretical performance ceiling. However, among the end-to-end compilation frameworks, \textit{nncase} demonstrates superior efficiency.

\textbf{Quantitative Analysis:}
\begin{itemize}
    \item \textbf{Qwen3-0.6B (F32):} \textit{nncase} achieves 8.7 tokens/s. Although this trails the hand-optimized \textit{llama.cpp} (10.61 tokens/s) by approximately 18\%, it notably outperforms \textit{Intel IPEX} (7.58 tokens/s) by \textbf{15\%}.
    \item \textbf{Qwen3-0.6B (F16):} Reducing precision significantly increases throughput. \textit{nncase} reaches 13.87 tokens/s, a \textbf{59\%} improvement over its F32 performance. This narrows the gap with \textit{llama.cpp} (17.21 tokens/s) and extends the lead over \textit{Intel IPEX} (10.22 tokens/s).
    \item \textbf{Qwen3-1.7B (F16):} As the model scale increases, the pressure of the memory bandwidth intensifies. \textit{nncase} maintains its ranking with 5.09 tokens/s, which is 19\% lower than \textit{llama.cpp} , but remains \textbf{21\% higher} than \textit{Intel IPEX}. \textit{MLC LLM} struggles in this configuration, achieving only 0.2 tokens/s.
\end{itemize}

These results validate that while there is still a marginal gap compared to heavily manually optimized libraries, \textit{nncase}' automated optimization pipeline generates significantly more efficient code than other state-of-the-art compilers (IPEX and MLC) in a strictly constrained single-core environment.

\vspace{-0.2cm}
\subsection{Multi-Core Scenario}
\label{subsec:multi_core}

\begin{figure}[t]
    \centering
    \includegraphics[width=\linewidth]{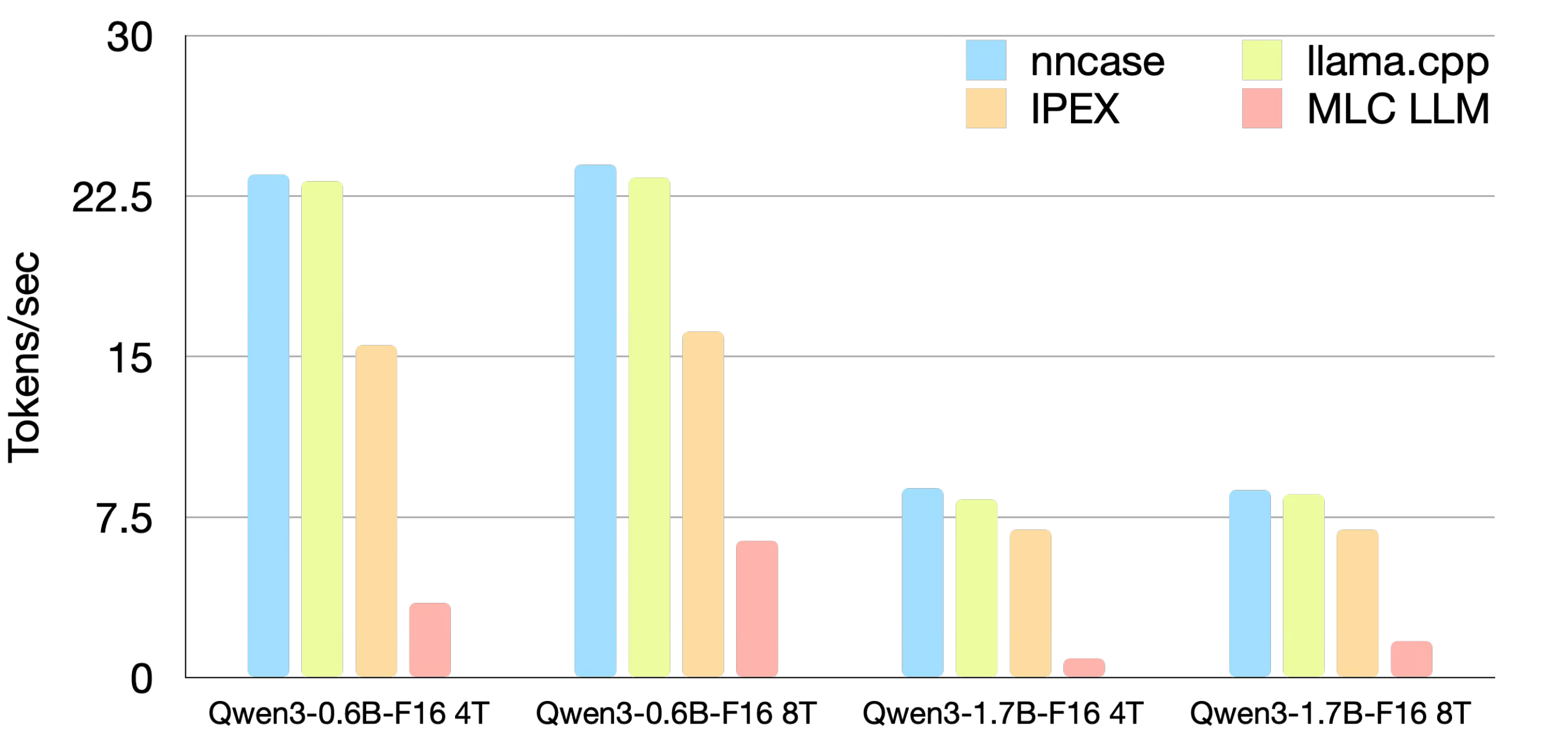}
    \caption{Comparison of LLM token generation throughput in multi-core scenarios (4T/8T). \textit{nncase} demonstrates superior scalability, surpassing hand-optimized libraries.}
    \vspace{-0.3cm}
    \label{fig:eval:multi_core}
\end{figure}

This subsection evaluates the efficacy of the \textit{Auto Distribution} module by scaling the workload to multi-core configurations (4T and 8T). Unlike traditional shared-memory parallelism, \textit{nncase} treats multiple cores as ``multi-nodes,'' applying distributed partition strategies to minimize synchronization overhead.

As shown in \autoref{fig:eval:multi_core}, \textit{nncase} exhibits the best scalability and overall performance, successfully overtaking the hand-optimized \textit{llama.cpp} in multi-core scenarios.

\textbf{Performance Analysis:}
\begin{itemize}
    \item \textbf{Scalability \& Throughput (Qwen3-0.6B-F16):} In the 4T configuration, \textit{nncase} achieves \textbf{23.5 tokens/s}, slightly outperforming \textit{llama.cpp} (23.2 tokens/s) and significantly leading \textit{Intel IPEX} (15.52 tokens/s). As the core count increases to 8T, the performance of all frameworks hits the memory bandwidth wall. Nevertheless, \textit{nncase} maintains the lead with 23.98 tokens/s.
    \item \textbf{Scaling Efficiency (Qwen3-1.7B-F16):} The advantage of \textit{nncase} becomes more pronounced with larger models. Comparing 4T with the 1T baseline (Sec.~\ref{subsec:single_core}), \textit{nncase} achieves a \textbf{performance gain} of 74\%, while \textit{llama.cpp} only gains \textbf{32\%}. This allows \textit{nncase} (8.85 tokens/s) to surpass \textit{llama.cpp} (8.34 tokens/s) and \textit{Intel IPEX} (6.93 tokens/s).
\end{itemize}

\textbf{Architectural Discussion:}
The superior multi-core performance of \textit{nncase} stems from its \textbf{Unified Distributed Compilation} paradigm. 
\begin{enumerate}
    \item \textbf{Static vs. Dynamic Scheduling:} Competitors like \textit{llama.cpp} and \textit{Intel IPEX} rely on OpenMP~\cite{openmp} for thread-level parallelism. This approach introduces runtime overhead due to dynamic task scheduling and frequent fork-join synchronization barriers. In contrast, \textit{nncase}' \textit{Auto Distribution} module performs static task partitioning and core mapping at compile time. This eliminates runtime scheduling overhead and ensures deterministic communication patterns.
    \item \textbf{Granularity Optimization:} By viewing cores as distributed nodes, the compiler optimizes the computation granularity to balance load and minimize inter-core communication, allowing for a near-linear release of parallel computing power before hitting hardware bandwidth limits.
\end{enumerate}

These results confirm that applying distributed system methodologies to single-machine multi-core compilation yields higher parallel efficiency than traditional shared-memory threading models.

\vspace{-0.3cm}

\section{Related Work}
\label{sec:related_work}

\subsection{Graph Rewriting Optimization}
Graph rewriting is crucial to enhancing execution efficiency in deep learning compilers. Existing approaches generally fall into two categories: \textit{rule-based superoptimization} and \textit{equality saturation}.

Rule-based methods optimize graphs through specific transformation rules and search strategies. TASO~\cite{taso} automates rule generation and verifies correctness through theorem provers, using cost-driven backtracking for search. Others, such as OCGGS~\cite{ocggs}, PET~\cite{pet}, and OLLIE~\cite{ollie}, extend this by relaxing performance constraints or deriving tensor algebra transformations. TPP~\cite{tpp_mlir} facilitates efficient vectorization through cache-aware packing primitives. However, these methods suffer from the \textit{phase ordering problem}, where the sequence of optimizations impacts the final result.

Equality saturation addresses phase ordering by maintaining equivalent program versions in an e-graph. Diospyros~\cite{diospyros} and Glenside~\cite{glenside} apply this to hardware adaptation, optimizing vectorization and access patterns. To improve the search efficiency of e-graphs, Tensat~\cite{tensat} and MCTS-GE~\cite{tensat_mcts} introduce techniques like Monte Carlo Tree Search.

\textbf{Limitations \& Our Approach:} Despite these advances, existing e-graph methods rarely support \textbf{layout optimization for heterogeneous computing units} or link data locality with \textbf{adaptive memory allocation}. Most treat layout and memory as post-optimization passes. \textit{nncase} bridges this gap by encoding layout and storage constraints directly into rewriting rules. By integrating a precise cost model within the e-graph, \textit{nncase} simultaneously optimizes algebraic structure, data layout, and buffer placement.

\subsection{Distribution Strategy Search}
Automatic distributed parallelism research focuses on strategy representation and search algorithms.
Regarding representation, OneFlow~\cite{oneflow} and AutoDDL~\cite{autoddl} utilize SBP abstractions (Split, Broadcast, Partial-Value) to manage resource dependencies. GSPMD~\cite{gspmd} supports hybrid parallelism but relies on manual annotations.
For search strategies, Alpa~\cite{alpa} constructs a hierarchical space for inter-operator and intra-operator parallelism, while Unity~\cite{unity} proposes joint optimization of algebraic transformations and parallelization. In complex scenarios, systems such as PaddleDist~\cite{paddle_dist} and Colossal-Auto~\cite{colossal-auto} address adaptive training and fault tolerance.

\textbf{Limitations \& Our Approach:} A critical limitation of current schemes is the \textbf{decoupling} of distributed strategy search from graph rewriting. Most systems (e.g., Alpa) perform the distribution search as an independent pass, preventing joint optimization with operator fusion or layout changes. \textit{nncase} overcomes this by reusing the e-graph to construct a unified search space. This allows distributed strategies to be co-optimized with graph rewriting without manual configuration, while explicitly modeling communication costs and storage constraints for heterogeneous hardware.

\subsection{Scheduling Optimization} 
Scheduling optimization balances code performance against search time. Approaches are broadly categorized into \textit{learning-based} methods and \textit{analytical modeling-based} methods.

Learning-based methods (e.g. AutoTVM~\cite{autotvm}, Ansor~\cite{ansor}, MetaSchedule~\cite{meta_schedule}) use machine learning to explore vast search spaces. Although they can approach optimal performance, they suffer from extremely long compilation times.
Analytical methods (e.g. Welder~\cite{welder}, TileFlow~\cite{tileflow}, Chimera~\cite{chimera}) use mathematical models to quickly solve parameters such as tile sizes. However, they often oversimplify the hardware model. For example, Tuna~\cite{tuna} is limited to scalar loops, and tools such as Welder often prioritize data movement over computational overhead, failing to accurately model the trade-off between cache size and memory footprint in complex hierarchies.

\textbf{Limitations \& Our Approach:} Existing analytical models often lack the precision to handle the complex constraints of specialized AI accelerators. \textit{nncase} addresses this by employing \textbf{MINLP (Mixed-Integer Non-Linear Programming)} modeling. Unlike prior works, \textit{nncase} explicitly characterizes buffer locations and incorporates precise computational overhead into the cost evaluation, enabling efficient generation of high-performance schedules with significantly reduced compilation time.

\vspace{-0.3cm}

\section{Conclusion and Future Work}
\label{sec:conclusion}

This paper introduces \textit{nncase}, an end-to-end compilation framework designed to overcome the challenges of deploying LLMs on heterogeneous storage architectures. 
By constructing a unified distributed compilation framework, \textit{nncase} bridges the gap between high-level model definitions and low-level hardware constraints. 
Key innovations include the e-graph-based Auto Vectorize and Auto Distribution modules, together with a hierarchical Auto Schedule module that synergizes MCTS with MINLP and integrates a specialized NTT library. 
These components collectively address critical bottlenecks in heterogeneous unit adaptation, distributed strategy search, and on-chip storage management. 
Experimental evaluations on modern multi-core platforms (e.g., AMD Ryzen 9 5900X) demonstrate that \textit{nncase} achieves superior inference performance for Qwen3 series models compared to state-of-the-art frameworks like Intel IPEX and MLC LLM, validating the efficacy of its unified design philosophy.

Looking ahead, we plan to advance this research in three strategic directions:
\begin{enumerate}
    \item \textbf{Holistic Performance Characterization:} We will expand the scope of evaluation to cover a wider range of model sizes, hardware platforms, and deployment scenarios. This includes a rigorous analysis of compilation overhead and memory footprint to verify the robustness and scalability of \textit{nncase} in complex production environments.
    
    \item \textbf{Cross-Architecture Adaptation (SIMT Support):} We aim to extend the compiler's backend to support SIMT architectures. By enhancing the Auto Vectorize and Auto Schedule modules, we will enable seamless deployment on massively parallel devices such as GPUs, achieving true cross-platform portability.
    
    \item \textbf{Computation-Communication Overlap:} We will focus on the automatic fusion and scheduling of communication and computation kernels. By refining dependency analysis and implementing advanced pipelining strategies, we aim to achieve effective latency hiding, thereby maximizing hardware utilization and end-to-end inference throughput in distributed settings.
\end{enumerate}

\clearpage
\bibliographystyle{plain}
\bibliography{main}

@inproceedings{autotvm,
author = {Chen, Tianqi and Zheng, Lianmin and Yan, Eddie and Jiang, Ziheng and Moreau, Thierry and Ceze, Luis and Guestrin, Carlos and Krishnamurthy, Arvind},
title = {Learning to optimize tensor programs},
year = {2018},
publisher = {Curran Associates Inc.},
address = {Red Hook, NY, USA},
booktitle = {Proceedings of the 32nd International Conference on Neural Information Processing Systems},
pages = {3393–3404},
numpages = {12},
location = {Montr\'{e}al, Canada},
series = {NIPS'18}
}

@misc{gspmd,
      title={GSPMD: General and Scalable Parallelization for ML Computation Graphs}, 
      author={Yuanzhong Xu and HyoukJoong Lee and Dehao Chen and Blake Hechtman and Yanping Huang and Rahul Joshi and Maxim Krikun and Dmitry Lepikhin and Andy Ly and Marcello Maggioni and Ruoming Pang and Noam Shazeer and Shibo Wang and Tao Wang and Yonghui Wu and Zhifeng Chen},
      year={2021},
      eprint={2105.04663},
      archivePrefix={arXiv},
      primaryClass={cs.DC},
      url={https://arxiv.org/abs/2105.04663}, 
}

@misc{paddle_dist,
      title={End-to-end Adaptive Distributed Training on PaddlePaddle}, 
      author={Yulong Ao and Zhihua Wu and Dianhai Yu and Weibao Gong and Zhiqing Kui and Minxu Zhang and Zilingfeng Ye and Liang Shen and Yanjun Ma and Tian Wu and Haifeng Wang and Wei Zeng and Chao Yang},
      year={2021},
      eprint={2112.02752},
      archivePrefix={arXiv},
      primaryClass={cs.DC},
      url={https://arxiv.org/abs/2112.02752}, 
}

@misc{colossal-auto,
      title={Colossal-Auto: Unified Automation of Parallelization and Activation Checkpoint for Large-scale Models}, 
      author={Yuliang Liu and Shenggui Li and Jiarui Fang and Yanjun Shao and Boyuan Yao and Yang You},
      year={2023},
      eprint={2302.02599},
      archivePrefix={arXiv},
      primaryClass={cs.LG},
      url={https://arxiv.org/abs/2302.02599}, 
}

@inproceedings{glenside,
author = {Smith, Gus Henry and Liu, Andrew and Lyubomirsky, Steven and Davidson, Scott and McMahan, Joseph and Taylor, Michael and Ceze, Luis and Tatlock, Zachary},
title = {Pure tensor program rewriting via access patterns (representation pearl)},
year = {2021},
isbn = {9781450384674},
publisher = {Association for Computing Machinery},
address = {New York, NY, USA},
url = {https://doi.org/10.1145/3460945.3464953},
doi = {10.1145/3460945.3464953},
booktitle = {Proceedings of the 5th ACM SIGPLAN International Symposium on Machine Programming},
pages = {21–31},
numpages = {11},
location = {Virtual, Canada},
series = {MAPS 2021}
}

@misc{ollie,
      title={OLLIE: Derivation-based Tensor Program Optimizer}, 
      author={Liyan Zheng and Haojie Wang and Jidong Zhai and Muyan Hu and Zixuan Ma and Tuowei Wang and Shizhi Tang and Lei Xie and Kezhao Huang and Zhihao Jia},
      year={2022},
      eprint={2208.02025},
      archivePrefix={arXiv},
      primaryClass={cs.LG},
      url={https://arxiv.org/abs/2208.02025}, 
}

@inproceedings{unity,
  title={Unity: Accelerating DNN Training Through Joint Optimization of Algebraic Transformations and Parallelization},
  author={Colin Unger and Zhihao Jia and Wei Wu and Sina Lin and Mandeep Baines and Carlos Efrain Quintero Narvaez and Vinay Ramakrishnaiah and Nirmal Prajapati and Patrick S. McCormick and Jamal Mohd-Yusof and Xi Luo and Dheevatsa Mudigere and Jongsoo Park and Misha Smelyanskiy and Alexander Aiken},
  booktitle={USENIX Symposium on Operating Systems Design and Implementation},
  year={2022},
  url={https://api.semanticscholar.org/CorpusID:267834645}
}

@inproceedings{ocggs,
 author = {Jia, Zhihao and Thomas, James and Warszawski, Todd and Gao, Mingyu and Zaharia, Matei and Aiken, Alex},
 booktitle = {Proceedings of Machine Learning and Systems},
 editor = {A. Talwalkar and V. Smith and M. Zaharia},
 pages = {27--39},
 title = {Optimizing DNN Computation with Relaxed Graph Substitutions},
 url = {https://proceedings.mlsys.org/paper_files/paper/2019/file/4dd1a7279a8cfeea2660fbc34f02a2bc-Paper.pdf},
 volume = {1},
 year = {2019}
}

@inproceedings {pet,
author = {Haojie Wang and Jidong Zhai and Mingyu Gao and Zixuan Ma and Shizhi Tang and Liyan Zheng and Yuanzhi Li and Kaiyuan Rong and Yuanyong Chen and Zhihao Jia},
title = {{PET}: Optimizing Tensor Programs with Partially Equivalent Transformations and Automated Corrections},
booktitle = {15th USENIX Symposium on Operating Systems Design and Implementation (OSDI 21)},
year = {2021},
isbn = {978-1-939133-22-9},
pages = {37--54},
url = {https://www.usenix.org/conference/osdi21/presentation/wang},
publisher = {USENIX Association},
month = jul
}

@inproceedings{tensat_mcts,
author = {Hartmann, Jakob and He, Guoliang and Yoneki, Eiko},
title = {Optimizing Tensor Computation Graphs with Equality Saturation and Monte Carlo Tree Search},
year = {2024},
isbn = {9798400706318},
publisher = {Association for Computing Machinery},
address = {New York, NY, USA},
url = {https://doi.org/10.1145/3656019.3689611},
doi = {10.1145/3656019.3689611},
booktitle = {Proceedings of the 2024 International Conference on Parallel Architectures and Compilation Techniques},
pages = {40–52},
numpages = {13},
location = {Long Beach, CA, USA},
series = {PACT '24}
}

@misc{tensat,
      title={Equality Saturation for Tensor Graph Superoptimization}, 
      author={Yichen Yang and Phitchaya Mangpo Phothilimtha and Yisu Remy Wang and Max Willsey and Sudip Roy and Jacques Pienaar},
      year={2021},
      eprint={2101.01332},
      archivePrefix={arXiv},
      primaryClass={cs.AI},
      url={https://arxiv.org/abs/2101.01332}, 
}

@inproceedings{taso,
author = {Jia, Zhihao and Padon, Oded and Thomas, James and Warszawski, Todd and Zaharia, Matei and Aiken, Alex},
title = {TASO: optimizing deep learning computation with automatic generation of graph substitutions},
year = {2019},
isbn = {9781450368735},
publisher = {Association for Computing Machinery},
address = {New York, NY, USA},
url = {https://doi.org/10.1145/3341301.3359630},
doi = {10.1145/3341301.3359630},
booktitle = {Proceedings of the 27th ACM Symposium on Operating Systems Principles},
pages = {47–62},
numpages = {16},
location = {Huntsville, Ontario, Canada},
series = {SOSP '19}
}

@inproceedings{diospyros,
author = {VanHattum, Alexa and Nigam, Rachit and Lee, Vincent T. and Bornholt, James and Sampson, Adrian},
title = {Vectorization for digital signal processors via equality saturation},
year = {2021},
isbn = {9781450383172},
publisher = {Association for Computing Machinery},
address = {New York, NY, USA},
url = {https://doi.org/10.1145/3445814.3446707},
doi = {10.1145/3445814.3446707},
booktitle = {Proceedings of the 26th ACM International Conference on Architectural Support for Programming Languages and Operating Systems},
pages = {874–886},
numpages = {13},
location = {Virtual, USA},
series = {ASPLOS '21}
}

@ARTICLE{openmp,
  author={Dagum, L. and Menon, R.},
  journal={IEEE Computational Science and Engineering}, 
  title={OpenMP: an industry standard API for shared-memory programming}, 
  year={1998},
  volume={5},
  number={1},
  pages={46-55},
  doi={10.1109/99.660313}}

@misc{qwen3,
      title={Qwen3 Technical Report}, 
      author={An Yang and Anfeng Li and Baosong Yang and Beichen Zhang and Binyuan Hui and Bo Zheng and Bowen Yu and Chang Gao and Chengen Huang and Chenxu Lv and Chujie Zheng and Dayiheng Liu and Fan Zhou and Fei Huang and Feng Hu and Hao Ge and Haoran Wei and Huan Lin and Jialong Tang and Jian Yang and Jianhong Tu and Jianwei Zhang and Jianxin Yang and Jiaxi Yang and Jing Zhou and Jingren Zhou and Junyang Lin and Kai Dang and Keqin Bao and Kexin Yang and Le Yu and Lianghao Deng and Mei Li and Mingfeng Xue and Mingze Li and Pei Zhang and Peng Wang and Qin Zhu and Rui Men and Ruize Gao and Shixuan Liu and Shuang Luo and Tianhao Li and Tianyi Tang and Wenbiao Yin and Xingzhang Ren and Xinyu Wang and Xinyu Zhang and Xuancheng Ren and Yang Fan and Yang Su and Yichang Zhang and Yinger Zhang and Yu Wan and Yuqiong Liu and Zekun Wang and Zeyu Cui and Zhenru Zhang and Zhipeng Zhou and Zihan Qiu},
      year={2025},
      eprint={2505.09388},
      archivePrefix={arXiv},
      primaryClass={cs.CL},
      url={https://arxiv.org/abs/2505.09388}, 
}

@software{mlc-llm,
    author = {{MLC team}},
    title = {{MLC-LLM}},
    url = {https://github.com/mlc-ai/mlc-llm},
    year = {2023-2025}
}

@misc{ipex,
  author = {intel},
  title = {Intel Extension for PyTorch},
  year = {2022},
  publisher = {GitHub},
  journal = {GitHub repository},
  howpublished = {\url{https://github.com/intel/intel-extension-for-pytorch}},
  commit = {cb81bf24b07c5b9eff726f4104062ee9d0d0246a}
}

@misc{llama_cpp,
  author = {Gerganov, Georgi},
  title = {llama.cpp},
  year = {2022},
  publisher = {GitHub},
  journal = {GitHub repository},
  howpublished = {\url{https://github.com/ggml-org/llama.cpp}},
  commit = {73e53dc834c0a2336cd104473af6897197b96277}
}

@incollection{binpack2,
  author       = {Edward G. Coffman Jr. and
                  G{\'{a}}bor Galambos and
                  Silvano Martello and
                  Daniele Vigo},
  editor       = {Ding{-}Zhu Du and
                  Panos M. Pardalos},
  title        = {Bin Packing Approximation Algorithms: Combinatorial Analysis},
  booktitle    = {Handbook of Combinatorial Optimization},
  pages        = {151--207},
  publisher    = {Springer},
  year         = {1999},
  url          = {https://doi.org/10.1007/978-1-4757-3023-4\_3},
  doi          = {10.1007/978-1-4757-3023-4\_3},
  bibsource    = {dblp computer science bibliography, https://dblp.org}
}

@article{binpack,
  author       = {Silvano Martello and
                  Paolo Toth},
  title        = {Lower bounds and reduction procedures for the bin packing problem},
  journal      = {Discret. Appl. Math.},
  volume       = {28},
  number       = {1},
  pages        = {59--70},
  year         = {1990},
  url          = {https://doi.org/10.1016/0166-218X(90)90094-S},
  doi          = {10.1016/0166-218X(90)90094-S},
  bibsource    = {dblp computer science bibliography, https://dblp.org}
}

@book{opt_book,
author = {Korte, Bernhard and Vygen, Jens},
title = {Combinatorial Optimization: Theory and Algorithms},
year = {2012},
isbn = {3642244874},
publisher = {Springer Publishing Company, Incorporated},
edition = {5th}
}

@article{mpich,
author = {Thakur, Rajeev and Rabenseifner, Rolf and Gropp, William},
year = {2005},
month = {01},
pages = {49-66},
title = {Optimization of Collective Communication Operations in MPICH.},
volume = {19},
journal = {IJHPCA}
}

@article{egraph_extract_sat, 
	author = {Mike He and Haichen Dong and Sharad Malik and Aarti Gupta}, 
	title = {Improving Term Extraction with Acyclic Constraints}, 
	booktitle = {E-Graph Research, Applications, Practices, and Human-factors Symposium (PLDI/EGRAPHS'23)}, 
    journal = "Programming Language Design and Implementation",
	year = {2023}, 
}

@inproceedings{equality_saturated_2,
author = {Tate, Ross and Stepp, Michael and Tatlock, Zachary and Lerner, Sorin},
title = {Equality saturation: a new approach to optimization},
year = {2009},
isbn = {9781605583792},
publisher = {Association for Computing Machinery},
address = {New York, NY, USA},
url = {https://doi.org/10.1145/1480881.1480915},
doi = {10.1145/1480881.1480915},
booktitle = {Proceedings of the 36th Annual ACM SIGPLAN-SIGACT Symposium on Principles of Programming Languages},
pages = {264–276},
numpages = {13},
location = {Savannah, GA, USA},
series = {POPL '09}
}

@inproceedings{equality_saturated_1,
author = {Stepp, Michael and Tate, Ross and Lerner, Sorin},
title = {Equality-based translation validator for LLVM},
year = {2011},
isbn = {9783642221095},
publisher = {Springer-Verlag},
address = {Berlin, Heidelberg},
booktitle = {Proceedings of the 23rd International Conference on Computer Aided Verification},
pages = {737–742},
numpages = {6},
location = {Snowbird, UT},
series = {CAV'11}
}

@inproceedings{phase_order_2,
author = {Kulkarni, Prasad and Zhao, Wankang and Moon, Hwashin and Cho, Kyunghwan and Whalley, David and Davidson, Jack and Bailey, Mark and Paek, Yunheung and Gallivan, Kyle},
title = {Finding effective optimization phase sequences},
year = {2003},
isbn = {1581136471},
publisher = {Association for Computing Machinery},
address = {New York, NY, USA},
url = {https://doi.org/10.1145/780732.780735},
doi = {10.1145/780732.780735},
booktitle = {Proceedings of the 2003 ACM SIGPLAN Conference on Language, Compiler, and Tool for Embedded Systems},
pages = {12–23},
numpages = {12},
location = {San Diego, California, USA},
series = {LCTES '03}
}

@inproceedings{phase_order_1,
author = {Benitez, M. E. and Davidson, J. W.},
title = {A portable global optimizer and linker},
year = {1988},
isbn = {0897912691},
publisher = {Association for Computing Machinery},
address = {New York, NY, USA},
url = {https://doi.org/10.1145/53990.54023},
doi = {10.1145/53990.54023},
booktitle = {Proceedings of the ACM SIGPLAN 1988 Conference on Programming Language Design and Implementation},
pages = {329–338},
numpages = {10},
location = {Atlanta, Georgia, USA},
series = {PLDI '88}
}

@inproceedings{term_rewrite,
author = {Dershowitz, Nachum},
title = {A Taste of Rewrite Systems},
year = {1993},
isbn = {3540568832},
publisher = {Springer-Verlag},
address = {Berlin, Heidelberg},
booktitle = {Functional Programming, Concurrency, Simulation and Automated Reasoning: International Lecture Series 1991-1992, McMaster University, Hamilton, Ontario, Canada},
pages = {199–228},
numpages = {30}
}

@inproceedings{tileflow,
author = {Zheng, Size and Chen, Siyuan and Gao, Siyuan and Jia, Liancheng and Sun, Guangyu and Wang, Runsheng and Liang, Yun},
title = {TileFlow: A Framework for Modeling Fusion Dataflow via Tree-based Analysis},
year = {2023},
isbn = {9798400703294},
publisher = {Association for Computing Machinery},
address = {New York, NY, USA},
url = {https://doi.org/10.1145/3613424.3623792},
doi = {10.1145/3613424.3623792},
booktitle = {Proceedings of the 56th Annual IEEE/ACM International Symposium on Microarchitecture},
pages = {1271–1288},
numpages = {18},
location = {Toronto, ON, Canada},
series = {MICRO '23}
}

@article{roofline,
author = {Williams, Samuel and Waterman, Andrew and Patterson, David},
title = {Roofline: an insightful visual performance model for multicore architectures},
year = {2009},
issue_date = {April 2009},
publisher = {Association for Computing Machinery},
address = {New York, NY, USA},
volume = {52},
number = {4},
issn = {0001-0782},
url = {https://doi.org/10.1145/1498765.1498785},
doi = {10.1145/1498765.1498785},
journal = {Commun. ACM},
month = apr,
pages = {65–76},
numpages = {12}
}

@INPROCEEDINGS{chimera,
  author={Zheng, Size and Chen, Siyuan and Song, Peidi and Chen, Renze and Li, Xiuhong and Yan, Shengen and Lin, Dahua and Leng, Jingwen and Liang, Yun},
  booktitle={2023 IEEE International Symposium on High-Performance Computer Architecture (HPCA)}, 
  title={Chimera: An Analytical Optimizing Framework for Effective Compute-intensive Operators Fusion}, 
  year={2023},
  volume={},
  number={},
  pages={1113-1126},
  doi={10.1109/HPCA56546.2023.10071018}}

@article{analytical_blis,
author = {Low, Tze Meng and Igual, Francisco D. and Smith, Tyler M. and Quintana-Orti, Enrique S.},
title = {Analytical Modeling Is Enough for High-Performance BLIS},
year = {2016},
issue_date = {June 2017},
publisher = {Association for Computing Machinery},
address = {New York, NY, USA},
volume = {43},
number = {2},
issn = {0098-3500},
url = {https://doi.org/10.1145/2925987},
doi = {10.1145/2925987},
journal = {ACM Trans. Math. Softw.},
month = aug,
articleno = {12},
numpages = {18}
}

@misc{tuna,
      title={Tuna: A Static Analysis Approach to Optimizing Deep Neural Networks}, 
      author={Yao Wang and Xingyu Zhou and Yanming Wang and Rui Li and Yong Wu and Vin Sharma},
      year={2021},
      eprint={2104.14641},
      archivePrefix={arXiv},
      primaryClass={cs.DC},
      url={https://arxiv.org/abs/2104.14641}, 
}

@inproceedings{analytical_model,
author = {Li, Rui and Sukumaran-Rajam, Aravind and Veras, Richard and Low, Tze Meng and Rastello, Fabrice and Rountev, Atanas and Sadayappan, P.},
title = {Analytical cache modeling and tilesize optimization for tensor contractions},
year = {2019},
isbn = {9781450362290},
publisher = {Association for Computing Machinery},
address = {New York, NY, USA},
url = {https://doi.org/10.1145/3295500.3356218},
doi = {10.1145/3295500.3356218},
booktitle = {Proceedings of the International Conference for High Performance Computing, Networking, Storage and Analysis},
articleno = {74},
numpages = {13},
location = {Denver, Colorado},
series = {SC '19}
}

@inproceedings{meta_schedule,
author = {Shao, Junru and Zhou, Xiyou and Feng, Siyuan and Hou, Bohan and Lai, Ruihang and Jin, Hongyi and Lin, Wuwei and Masuda, Masahiro and Yu, Cody Hao and Chen, Tianqi},
title = {Tensor program optimization with probabilistic programs},
year = {2022},
isbn = {9781713871088},
publisher = {Curran Associates Inc.},
address = {Red Hook, NY, USA},
booktitle = {Proceedings of the 36th International Conference on Neural Information Processing Systems},
articleno = {2593},
numpages = {14},
location = {New Orleans, LA, USA},
series = {NIPS '22}
}

@inproceedings{ansor,
author = {Zheng, Lianmin and Jia, Chengfan and Sun, Minmin and Wu, Zhao and Yu, Cody Hao and Haj-Ali, Ameer and Wang, Yida and Yang, Jun and Zhuo, Danyang and Sen, Koushik and Gonzalez, Joseph E. and Stoica, Ion},
title = {Ansor: generating high-performance tensor programs for deep learning},
year = {2020},
isbn = {978-1-939133-19-9},
publisher = {USENIX Association},
address = {USA},
booktitle = {Proceedings of the 14th USENIX Conference on Operating Systems Design and Implementation},
articleno = {49},
numpages = {17},
series = {OSDI'20}
}

@article{auto_halide,
author = {Adams, Andrew and Ma, Karima and Anderson, Luke and Baghdadi, Riyadh and Li, Tzu-Mao and Gharbi, Micha\"{e}l and Steiner, Benoit and Johnson, Steven and Fatahalian, Kayvon and Durand, Fr\'{e}do and Ragan-Kelley, Jonathan},
title = {Learning to optimize halide with tree search and random programs},
year = {2019},
issue_date = {August 2019},
publisher = {Association for Computing Machinery},
address = {New York, NY, USA},
volume = {38},
number = {4},
issn = {0730-0301},
url = {https://doi.org/10.1145/3306346.3322967},
doi = {10.1145/3306346.3322967},
journal = {ACM Trans. Graph.},
month = jul,
articleno = {121},
numpages = {12}
}

@misc{looptune,
      title={LoopTune: Optimizing Tensor Computations with Reinforcement Learning}, 
      author={Dejan Grubisic and Bram Wasti and Chris Cummins and John Mellor-Crummey and Aleksandar Zlateski},
      year={2023},
      eprint={2309.01825},
      archivePrefix={arXiv},
      primaryClass={cs.LG},
      url={https://arxiv.org/abs/2309.01825}, 
}

@misc{looper,
      title={LOOPer: A Learned Automatic Code Optimizer For Polyhedral Compilers}, 
      author={Massinissa Merouani and Khaled Afif Boudaoud and Iheb Nassim Aouadj and Nassim Tchoulak and Islem Kara Bernou and Hamza Benyamina and Fatima Benbouzid-Si Tayeb and Karima Benatchba and Hugh Leather and Riyadh Baghdadi},
      year={2025},
      eprint={2403.11522},
      archivePrefix={arXiv},
      primaryClass={cs.PL},
      url={https://arxiv.org/abs/2403.11522}, 
}

@article{autoddl,
author = {Chen, Jinfan and Li, Shigang and Guo, Ran and Yuan, Jinhui and Hoefler, Torsten},
title = {AutoDDL: Automatic Distributed Deep Learning With Near-Optimal Bandwidth Cost},
year = {2024},
issue_date = {Aug. 2024},
publisher = {IEEE Press},
volume = {35},
number = {8},
issn = {1045-9219},
url = {https://doi.org/10.1109/TPDS.2024.3397800},
doi = {10.1109/TPDS.2024.3397800},
journal = {IEEE Trans. Parallel Distrib. Syst.},
month = aug,
pages = {1331–1344},
numpages = {14}
}

@inproceedings{alt,
author = {Xu, Zhiying and Xu, Jiafan and Peng, Hongding and Wang, Wei and Wang, Xiaoliang and Wan, Haoran and Dai, Haipeng and Xu, Yixu and Cheng, Hao and Wang, Kun and Chen, Guihai},
title = {ALT: Breaking the Wall between Data Layout and Loop Optimizations for Deep Learning Compilation},
year = {2023},
isbn = {9781450394871},
publisher = {Association for Computing Machinery},
address = {New York, NY, USA},
url = {https://doi.org/10.1145/3552326.3587440},
doi = {10.1145/3552326.3587440},
booktitle = {Proceedings of the Eighteenth European Conference on Computer Systems},
pages = {199–214},
numpages = {16},
location = {Rome, Italy},
series = {EuroSys '23}
}

@misc{tpp_mlir,
      title={Towards a high-performance AI compiler with upstream MLIR}, 
      author={Renato Golin and Lorenzo Chelini and Adam Siemieniuk and Kavitha Madhu and Niranjan Hasabnis and Hans Pabst and Evangelos Georganas and Alexander Heinecke},
      year={2024},
      eprint={2404.15204},
      archivePrefix={arXiv},
      primaryClass={cs.PL},
      url={https://arxiv.org/abs/2404.15204}, 
}

@article{gotoblas,
author = {Goto, Kazushige and Geijn, Robert A. van de},
title = {Anatomy of high-performance matrix multiplication},
year = {2008},
issue_date = {May 2008},
publisher = {Association for Computing Machinery},
address = {New York, NY, USA},
volume = {34},
number = {3},
issn = {0098-3500},
url = {https://doi.org/10.1145/1356052.1356053},
doi = {10.1145/1356052.1356053},
journal = {ACM Trans. Math. Softw.},
month = may,
articleno = {12},
numpages = {25}
}

@article{halide,
author = {Ragan-Kelley, Jonathan and Barnes, Connelly and Adams, Andrew and Paris, Sylvain and Durand, Fr\'{e}do and Amarasinghe, Saman},
title = {Halide: a language and compiler for optimizing parallelism, locality, and recomputation in image processing pipelines},
year = {2013},
issue_date = {June 2013},
publisher = {Association for Computing Machinery},
address = {New York, NY, USA},
volume = {48},
number = {6},
issn = {0362-1340},
url = {https://doi.org/10.1145/2499370.2462176},
doi = {10.1145/2499370.2462176},
journal = {SIGPLAN Not.},
month = jun,
pages = {519–530},
numpages = {12}
}

@manual{arm_acl,
    organization = "ARM Ltd",
    title = "Arm Compute Library Reference Manual"
}

@manual{intel_mkl,
    title =  "Intel Math Kernel Library Reference Manual",
    organization = "Intel Corporation"
}

@INPROCEEDINGS{apple_amx,
  author={Wilkinson, Finn and McIntosh-Smith, Simon},
  booktitle={2022 IEEE/ACM International Workshop on Performance Modeling, Benchmarking and Simulation of High Performance Computer Systems (PMBS)}, 
  title={An Initial Evaluation of Arm’s Scalable Matrix Extension}, 
  year={2022},
  volume={},
  number={},
  pages={135-140},
  doi={10.1109/PMBS56514.2022.00018}}

@book{armv8,
title = "Technology Preview: The ARMv8 Architecture",
author = "John Goodacre",
note = "White paper",
year = "2011",
month = nov,
language = "English",
publisher = "ARM Ltd",
address = "United Kingdom",
}

@techreport{arm_sme,
    author = "Martin Weidman",
    title = "Introducing the scalable matrix extension for the armv9-a architecture",
    institution = "ARM Ltd",
    year = "2021",
    url = "https://web.archive.org/web/20240710075527/https://community.arm.com/arm-community-blogs/b/architectures-and-processors-blog/posts/scalable-matrix-extension-armv9-a-architecture"
}

@misc{ortools,
  author = {},
  title = {or-tools},
  year = {2013},
  publisher = {GitHub},
  journal = {GitHub repository},
  howpublished = {\url{https://github.com/google/or-tools}},
  commit = {8d9216e14c7c84a875ccd7159fe4ea7d4d6434d5}
}

@article{branch-bounds-minlp,
author = {Belotti, Pietro and Lee, Jon and Liberti, Leo and Margot, Francois and Wachter, Andreas},
title = {Branching and bounds tighteningtechniques for non-convex MINLP},
year = {2009},
issue_date = {August 2009},
publisher = {Taylor \& Francis, Inc.},
address = {USA},
volume = {24},
number = {4–5},
issn = {1055-6788},
url = {https://doi.org/10.1080/10556780903087124},
doi = {10.1080/10556780903087124},
journal = {Optimization Methods Software},
month = aug,
pages = {597–634},
numpages = {38}
}

@article{mcts,
author = {Nicholas Metropolis and S. Ulam},
title = {The Monte Carlo Method},
journal = {Journal of the American Statistical Association},
volume = {44},
number = {247},
pages = {335--341},
year = {1949},
publisher = {ASA Website},
doi = {10.1080/01621459.1949.10483310},
note = {PMID: 18139350},
URL = {https://www.tandfonline.com/doi/abs/10.1080/01621459.1949.10483310},
eprint = {https://www.tandfonline.com/doi/pdf/10.1080/01621459.1949.10483310}
}

@misc{oneflow,
      title={OneFlow: Redesign the Distributed Deep Learning Framework from Scratch}, 
      author={Jinhui Yuan and Xinqi Li and Cheng Cheng and Juncheng Liu and Ran Guo and Shenghang Cai and Chi Yao and Fei Yang and Xiaodong Yi and Chuan Wu and Haoran Zhang and Jie Zhao},
      year={2022},
      eprint={2110.15032},
      archivePrefix={arXiv},
      primaryClass={cs.DC},
      url={https://arxiv.org/abs/2110.15032}, 
}

@ARTICLE{intel-xeonphi,
  author={Sodani, Avinash and Gramunt, Roger and Corbal, Jesus and Kim, Ho-Seop and Vinod, Krishna and Chinthamani, Sundaram and Hutsell, Steven and Agarwal, Rajat and Liu, Yen-Chen},
  journal={IEEE Micro}, 
  title={Knights Landing: Second-Generation Intel Xeon Phi Product}, 
  year={2016},
  volume={36},
  number={2},
  pages={34-46},
  doi={10.1109/MM.2016.25}}

@INPROCEEDINGS{intel-Xeon6,
  author={Varada, Raj R. and Krishnan, Rohini and Subramonia, Ajith and Chandran, Rathish and Chakravarthy, Kalyana and Desai, Uttpal D. and Limaye, Sumedha and Puri, Puneesh and Mulvihill, David R. and Bichan, Mike and Koolhaas, Martin and Ramachandran, Vijayalakshmi and Kendle, Srinivasu},
  booktitle={2025 IEEE International Solid-State Circuits Conference (ISSCC)}, 
  title={2.3 Granite Rapids-D: Intel Xeon 6 SoC for vRAN, Edge, Networking, and Storage}, 
  year={2025},
  volume={68},
  number={},
  pages={48-50},
  doi={10.1109/ISSCC49661.2025.10904676}}

@misc{llama,
      title={LLaMA: Open and Efficient Foundation Language Models}, 
      author={Hugo Touvron and Thibaut Lavril and Gautier Izacard and Xavier Martinet and Marie-Anne Lachaux and Timothée Lacroix and Baptiste Rozière and Naman Goyal and Eric Hambro and Faisal Azhar and Aurelien Rodriguez and Armand Joulin and Edouard Grave and Guillaume Lample},
      year={2023},
      eprint={2302.13971},
      archivePrefix={arXiv},
      primaryClass={cs.CL},
      url={https://arxiv.org/abs/2302.13971}, 
}

@misc{gpt4,
      title={GPT-4 Technical Report}, 
      author={OpenAI and Josh Achiam and Steven Adler and Sandhini Agarwal and Lama Ahmad and Ilge Akkaya and Florencia Leoni Aleman and Diogo Almeida and Janko Altenschmidt and Sam Altman and Shyamal Anadkat and Red Avila and Igor Babuschkin and Suchir Balaji and Valerie Balcom and Paul Baltescu and Haiming Bao and Mohammad Bavarian and Jeff Belgum and Irwan Bello and Jake Berdine and Gabriel Bernadett-Shapiro and Christopher Berner and Lenny Bogdonoff and Oleg Boiko and Madelaine Boyd and Anna-Luisa Brakman and Greg Brockman and Tim Brooks and Miles Brundage and Kevin Button and Trevor Cai and Rosie Campbell and Andrew Cann and Brittany Carey and Chelsea Carlson and Rory Carmichael and Brooke Chan and Che Chang and Fotis Chantzis and Derek Chen and Sully Chen and Ruby Chen and Jason Chen and Mark Chen and Ben Chess and Chester Cho and Casey Chu and Hyung Won Chung and Dave Cummings and Jeremiah Currier and Yunxing Dai and Cory Decareaux and Thomas Degry and Noah Deutsch and Damien Deville and Arka Dhar and David Dohan and Steve Dowling and Sheila Dunning and Adrien Ecoffet and Atty Eleti and Tyna Eloundou and David Farhi and Liam Fedus and Niko Felix and Simón Posada Fishman and Juston Forte and Isabella Fulford and Leo Gao and Elie Georges and Christian Gibson and Vik Goel and Tarun Gogineni and Gabriel Goh and Rapha Gontijo-Lopes and Jonathan Gordon and Morgan Grafstein and Scott Gray and Ryan Greene and Joshua Gross and Shixiang Shane Gu and Yufei Guo and Chris Hallacy and Jesse Han and Jeff Harris and Yuchen He and Mike Heaton and Johannes Heidecke and Chris Hesse and Alan Hickey and Wade Hickey and Peter Hoeschele and Brandon Houghton and Kenny Hsu and Shengli Hu and Xin Hu and Joost Huizinga and Shantanu Jain and Shawn Jain and Joanne Jang and Angela Jiang and Roger Jiang and Haozhun Jin and Denny Jin and Shino Jomoto and Billie Jonn and Heewoo Jun and Tomer Kaftan and Łukasz Kaiser and Ali Kamali and Ingmar Kanitscheider and Nitish Shirish Keskar and Tabarak Khan and Logan Kilpatrick and Jong Wook Kim and Christina Kim and Yongjik Kim and Jan Hendrik Kirchner and Jamie Kiros and Matt Knight and Daniel Kokotajlo and Łukasz Kondraciuk and Andrew Kondrich and Aris Konstantinidis and Kyle Kosic and Gretchen Krueger and Vishal Kuo and Michael Lampe and Ikai Lan and Teddy Lee and Jan Leike and Jade Leung and Daniel Levy and Chak Ming Li and Rachel Lim and Molly Lin and Stephanie Lin and Mateusz Litwin and Theresa Lopez and Ryan Lowe and Patricia Lue and Anna Makanju and Kim Malfacini and Sam Manning and Todor Markov and Yaniv Markovski and Bianca Martin and Katie Mayer and Andrew Mayne and Bob McGrew and Scott Mayer McKinney and Christine McLeavey and Paul McMillan and Jake McNeil and David Medina and Aalok Mehta and Jacob Menick and Luke Metz and Andrey Mishchenko and Pamela Mishkin and Vinnie Monaco and Evan Morikawa and Daniel Mossing and Tong Mu and Mira Murati and Oleg Murk and David Mély and Ashvin Nair and Reiichiro Nakano and Rajeev Nayak and Arvind Neelakantan and Richard Ngo and Hyeonwoo Noh and Long Ouyang and Cullen O'Keefe and Jakub Pachocki and Alex Paino and Joe Palermo and Ashley Pantuliano and Giambattista Parascandolo and Joel Parish and Emy Parparita and Alex Passos and Mikhail Pavlov and Andrew Peng and Adam Perelman and Filipe de Avila Belbute Peres and Michael Petrov and Henrique Ponde de Oliveira Pinto and Michael and Pokorny and Michelle Pokrass and Vitchyr H. Pong and Tolly Powell and Alethea Power and Boris Power and Elizabeth Proehl and Raul Puri and Alec Radford and Jack Rae and Aditya Ramesh and Cameron Raymond and Francis Real and Kendra Rimbach and Carl Ross and Bob Rotsted and Henri Roussez and Nick Ryder and Mario Saltarelli and Ted Sanders and Shibani Santurkar and Girish Sastry and Heather Schmidt and David Schnurr and John Schulman and Daniel Selsam and Kyla Sheppard and Toki Sherbakov and Jessica Shieh and Sarah Shoker and Pranav Shyam and Szymon Sidor and Eric Sigler and Maddie Simens and Jordan Sitkin and Katarina Slama and Ian Sohl and Benjamin Sokolowsky and Yang Song and Natalie Staudacher and Felipe Petroski Such and Natalie Summers and Ilya Sutskever and Jie Tang and Nikolas Tezak and Madeleine B. Thompson and Phil Tillet and Amin Tootoonchian and Elizabeth Tseng and Preston Tuggle and Nick Turley and Jerry Tworek and Juan Felipe Cerón Uribe and Andrea Vallone and Arun Vijayvergiya and Chelsea Voss and Carroll Wainwright and Justin Jay Wang and Alvin Wang and Ben Wang and Jonathan Ward and Jason Wei and CJ Weinmann and Akila Welihinda and Peter Welinder and Jiayi Weng and Lilian Weng and Matt Wiethoff and Dave Willner and Clemens Winter and Samuel Wolrich and Hannah Wong and Lauren Workman and Sherwin Wu and Jeff Wu and Michael Wu and Kai Xiao and Tao Xu and Sarah Yoo and Kevin Yu and Qiming Yuan and Wojciech Zaremba and Rowan Zellers and Chong Zhang and Marvin Zhang and Shengjia Zhao and Tianhao Zheng and Juntang Zhuang and William Zhuk and Barret Zoph},
      year={2024},
      eprint={2303.08774},
      archivePrefix={arXiv},
      primaryClass={cs.CL},
      url={https://arxiv.org/abs/2303.08774}, 
}

@inproceedings{ladder,
    author = {Wang, Lei and Ma, Lingxiao and Cao, Shijie and Zhang, Quanlu and Xue, Jilong and Shi, Yining and Zheng, Ningxin and Miao, Ziming and Yang, Fan and Cao, Ting and others},
    title = {Ladder: Enabling Efficient Low-Precision Deep Learning Computing through Hardware-aware Tensor Transformation},
    booktitle = {18th USENIX Symposium on Operating Systems Design and Implementation (OSDI 24)},
    pages = {307--323},
    year = {2024}
}

@online{wse3,
    author = {{Cerebras Systems}},
    title = {Third Generation 5nm {Wafer Scale Engine} ({WSE-3}) Powers Industry's Most Scalable {AI} Supercomputers},
    year = {2024},
    url = {https://cerebras.ai/press-release/cerebras-announces-third-generation-wafer-scale-engine},
    note = {Press release, March 13, 2024}
}

@article{wse,
    author = {Lie, Sean},
    title = {{Cerebras} Architecture Deep Dive: First Look Inside the Hardware/Software Co-Design for Deep Learning},
    journal = {IEEE Micro},
    volume = {43},
    number = {3},
    year = {2023},
    pages = {18--30}
}

@article{tvm,
    author = {Chen, T. and others},
    title = {{TVM}: An Automated End-to-End Optimization Stack for Deep Learning},
    journal = {SSP 2018},
    year = {2018}
}

@inproceedings{roller,
    author = {Zhu, Hongyu and Wu, Ruofan and Diao, Yijia and Ke, Shanbin and Li, Haoyu and Zhang, Chen and Xue, Jilong and Ma, Lingxiao and Xia, Yuqing and Cui, Wei and Yang, Fan and Yang, Mao and Zhou, Lidong and Cidon, Asaf and Pekhimenko, Gennady},
    title = {{ROLLER}: Fast and Efficient Tensor Compilation for Deep Learning},
    booktitle = {16th USENIX Symposium on Operating Systems Design and Implementation (OSDI 22)},
    year = {2022},
    pages = {233--248}
}

@inproceedings{welder,
    author = {Shi, Yining and Yang, Zhi and Xue, Jilong and Ma, Lingxiao and Xia, Yuqing and Miao, Ziming and Guo, Yuxiao and Yang, Fan and Zhou, Lidong},
    title = {{Welder}: Scheduling Deep Learning Memory Access via Tile-graph},
    booktitle = {17th USENIX Symposium on Operating Systems Design and Implementation (OSDI 23)},
    year = {2023},
    pages = {701--718}
}

@inproceedings{alpa,
    author = {Zheng, Lianmin and Li, Zhuohan and Zhang, Hao and Zhuang, Yonghao and Chen, Zhifeng and Huang, Yanping and Wang, Yida and Xu, Yuanzhong and Zhuo, Danyang and Xing, Eric P and others},
    title = {{Alpa}: Automating inter-and intra-operator parallelism for distributed deep learning},
    booktitle = {16th USENIX Symposium on Operating Systems Design and Implementation (OSDI 22)},
    year = {2022},
    pages = {559--578}
}


\end{document}